\documentclass[a4paper,twoside]{article}

\usepackage{epsfig}
\usepackage{subcaption}
\usepackage{calc}
\usepackage{amssymb}
\usepackage{amstext}
\usepackage{amsmath}
\usepackage{amsthm}
\usepackage{multicol}
\usepackage{pslatex}
\usepackage{apalike}

\usepackage{algorithm2e}
\SetKwFunction{clamp}{clamp}
\SetKwFunction{newD}{D}
\SetKwFunction{lerp}{lerp}
\SetKwFunction{min}{min}
\SetKwFunction{max}{max}
\SetKwFunction{smoothstep}{smoothstep}
\SetKwFunction{sampleAlpha}{sampleAlpha}
\SetKwFunction{saturate}{saturate}

\usepackage{color,soul}

\usepackage{multirow}

\usepackage{float}

\usepackage{hyperref}

\let\bibhang\relax
\usepackage{natbib}

\usepackage{SCITEPRESS}     

\begin{document}

\title{A Hybrid System for Real-Time Rendering of Depth of Field Effect in Games}

\author{\authorname{Yu Wei Tan\sup{1}\orcidAuthor{0000-0002-7972-2828}, Nicholas Chua\sup{1}, Nathan Biette\sup{1}\orcidAuthor{0000-0001-7827-1538} and Anand Bhojan\sup{1}\orcidAuthor{0000-0001-8105-1739}}
\affiliation{\sup{1}School of Computing, National University of Singapore}
\email{\{yuwei, nicholaschuayunzhi, nathan.biette\}@u.nus.edu, banand@comp.nus.edu.sg}
}


\abstract{Real-time depth of field in game cinematics tends to approximate the semi-transparent silhouettes of out-of-focus objects through post-processing techniques. We leverage ray tracing hardware acceleration and spatio-temporal reconstruction to improve the realism of such semi-transparent regions through hybrid rendering, while maintaining interactive frame rates for immersive gaming. This paper extends our previous work with a complete presentation of our technique and details on its design, implementation, and future work.}

\onecolumn \maketitle \normalsize \setcounter{footnote}{0} \vfill

\section{\uppercase{Introduction}}

We present the design and evaluation of a novel real-time hybrid rendering approach for the Depth of Field (DoF) effect which incorporates post-process based DoF with temporally and spatially reconstructed ray trace based DoF. By adaptively combining the output of different passes, we achieve more accurate semi-transparencies of foreground geometry to reveal background objects. We believe that our hybrid DoF technique is the first to integrate a ray-traced output with a traditional post-processing pipeline. 

Building on our previous work \citep{Tan:2020:HDF}, the key contributions of this paper are as follows.
\begin{itemize}
	\item Design and implementation of a real-time hybrid rendering pipeline for DoF.  
	\item Visual quality evaluation of the hybrid method, specifically, the accuracy of semi-transparencies.
	\item Performance evaluation and trade-offs in the use of ray tracing for DoF.
\end{itemize}

\subsection{Background Information}

Current DoF implementations in game engines typically use the thin lens model \citep{Potmesil:1982:SIG} to approximate the behaviour of cameras. The zone of focus is the part of the scene where the objects look sharp. The Circle of Confusion (CoC) \citep{Demers:2004:DFS} of points in the zone of focus are smaller than a cell on the sensor, yielding a single pixel in the image, whereas points outside the zone of focus appear as a spot on the image based on their CoC. For such points which lie on the same object, an overall blur of the object is produced. 

Bokeh shapes, which are bright spots created by a beam of unfocused light hitting the camera sensor, appear in areas out of the zone of focus. They usually take the shape of the camera's aperture and can have circular or polygonal frames depending on the number of blades in the camera shutter. 

Blurred foreground objects also have a slightly transparent silhouette through which background colour can be observed. These semi-transparent edges cannot be properly rendered in games with post-processing as the image does not store any information behind a foreground object \citep{Kraus:2007:DOF}. However, such approaches are widely used in real-time rendering as images produced by rasterization are in sharp focus \citep{McGraw:2015:FBE}. According to \citet{Jimenez:2014:ARR}, many techniques can only perform an approximation of the background colour locally using neighbouring pixels like in \citet{Abadie:2018:ARR} or grow blur out of the silhouette of foreground objects onto background colour, reusing foreground information to avoid reconstructing the missing background. However, shifting the blur outwards from foreground objects produces inaccuracies with regards to their actual geometries, especially when the amount of extended area is comparable to the size of the objects themselves. Objects with more elaborate shapes also become fat and deformed at areas with large CoC. Nonetheless, such inaccuracies do not exist in ray-traced DoF \citep{Cook:1984:DRT} as we can simulate a thin lens and query the scene for intersections, not being limited to what is rendered in the rasterized image. Nonetheless, achieving interactive frame rates with ray tracing is difficult due to the high computational costs of calculating ray-geometry intersections and multiple shading for each pixel, even with the latest GPUs developed for ray tracing. Hence, hybrid rendering, which aims to combine existing rasterization techniques with ray tracing, is being researched. 
\section{\uppercase{Related Work}}


\subsection{Hybrid Rendering}

Examples of hybrid rendering on related effects include \citet{Macedo:2018:FRR} and \citet{Marrs:2018:ATA} which invoke ray tracing for reflections and anti-aliasing respectively only on pixels where rasterization techniques are unable to achieve realistic or desirable results. \citet{Beck:1981:HGR}, \citet{Hertel:2009:HGR} and \citet{Lau:2009:FHS} employ the same strategy to produce accurate shadows.

The concept of hybrid rendering can also be extended to general rendering pipelines. For example, \citet{Cabeleira:2010:CRR} uses rasterization for diffuse illumination and ray tracing for reflections and refractions. \citet{Barre:2018:HRR} is also one such pipeline that has replaced effects like screen-space reflections with their ray trace counterparts to achieve better image quality. Another commonly-used approach is \citet{Chen:2007:UHZ}, the substitution of primary ray generation with rasterization in recursive ray tracing by \citet{Whitted:1979:IIM}. \citet{Andrade:2014:THB} improves upon this technique by observing a render time limit through the prioritization of only the most important scene objects for ray tracing.





\subsection{DoF}

Many DoF rendering techniques have been devised over the years. \citet{Potmesil:1982:SIG} first introduced the concept of CoC for a point based on a thin lens model which simulates the effects of the lens and aperture of a physical camera. It employs a post-processing technique that converts sampled points into their CoCs. The intensity distributions of CoCs overlapping with each pixel are then accumulated to produce the final colour for the pixel. \citet{Haeberli:1990:ABH} integrates images rendered from different sample points across the aperture of the lens with an accumulation buffer. On the other hand, \citet{Cook:1984:DRT} traces multiple rays from these different sample points on the lens into the scene using a technique now commonly known as distributed ray tracing, for which improvements in ray budget have been made in \citet{Hou:2010:MRT} and \citet{Lei:2013:ADF}. 

For rendering with real-time performance constraints, spatial reconstruction and temporal accumulation approaches have also been developed. For instance, \citet{Dayal:2005:AFR} introduces adaptive spatio-temporal sampling, choosing to sample more based on colour variance in the rendered image with selective rendering by \citet{Chalmers:2006:SRC} and favouring newer samples for temporal accumulation in dynamic scenes. \citet{Schied:2017:SVF} also uses temporal accumulation to raise the effective sample count on top of image reconstruction guided by variance estimation. Such techniques have been applied for DoF such as in \citet{Hach:2015:CBR}, \citet{Leimkuhler:2018:LKS}, \citet{Weier:2018:FDF}, \citet{Yan:2016:F4S} and \citet{Zhang:2019:SDL}. More advanced reconstruction techniques for DoF have also been introduced, such as \cite{Belcour:2013:5CT}, \citet{Lehtinen:2011:TLF}, \citet{Mehta:2014:FAF} and \citet{Vaidyanathan:2015:LLF} which sample light fields as well as \citet{Shirley:2011:LIR} which selectively blurs pixels of low frequency content in stochastic sampling. A more adaptive temporal accumulation approach from \citet{Schied:2018:GER} which is responsive to changes in sample attributes such as position and normal has also been proposed to mitigate ghosting and lag in classic temporal accumulation approaches.

Micropolygon-based techniques have also proven to be capable of DoF like in \citet{Fatahalian:2009:DRM} and \citet{Sattlecker:2015:RRG}. \citet{Catmull:1984:AVS} solves for per-pixel visibility by performing depth sorting on overlapping polygons for each pixel. Following this, approaches based on multi-layer images like \citet{Franke:2018:MDF}, \citet{Kraus:2007:DOF}, \citet{Lee:2008:RDR}, \citet{Lee:2009:DRM} and \citet{Selgrad:2015:RDF} have also been introduced where the contributions from each layer are accumulated to produce the final image. Such layered approaches are computationally expensive although they can generate relatively accurate results in terms of semi-transparencies. \citet{Bukowski:2013:SSF}, \citet{Jimenez:2014:ARR}, \citet{Valient:2013:KSF} and state-of-the-art Unreal Engine approach \citet{Abadie:2018:ARR} divide the scene into the background and foreground, and runs a gathering filter separately for each. We adopt such a technique, which performs better in terms of rendering time even in comparison to \citet{Yan:2016:F4S}, which avoids the problem of separating the scene by depth by factoring high-dimensional filters into 1D integrals.

\citet{Hach:2015:CBR} acquires a rich lens archive derived from a real target cinematic lens and uses it to synthesize a point spread function (PSF) for convolution in blurring. For each pixel, \citet{Leimkuhler:2018:LKS} splats its PSF using a sparse representation of its Laplacian. Time-dependent edge functions for \citet{Akenine:2007:SRT} and complex plane phasors for \citet{Garcia:2017:CSC} have also been used to produce DoF. Such approaches involve complex computations and seem to be more suitable for offline rendering. More recently, convolutional neural network approaches like \citet{Zhang:2019:SDL} perform post-processing for DoF by predicting the amount of blur to generate through the analysis of past frames in real-time but require copious amounts of training data.

\citet{McGraw:2015:FBE} and \citet{McIntosh:2012:ESB} are post-process techniques that produce polygonal bokeh shapes based on the silhouette of the camera aperture. \citeauthor{McGraw:2015:FBE} also supports bokeh shapes of non-uniform intensities, including bokeh shapes which are lighter or darker at the rim due to spherical aberration of the lens. Our approach currently generates circular bokeh shapes of uniform intensities but can be extended to produce alternative bokeh shapes such as polygons by changing the shape of our sampling kernel, and bokeh shapes of varying intensities by adjusting the relative weight of samples within the kernel.

Our hybrid DoF technique is novel as we augment conventional post-process approaches with ray tracing, generating more accurate semi-transparencies of foreground geometry in real-time.

\section{\uppercase{Design}}

\begin{figure*}[!h]
    \centering
    \includegraphics[width=\linewidth,keepaspectratio]{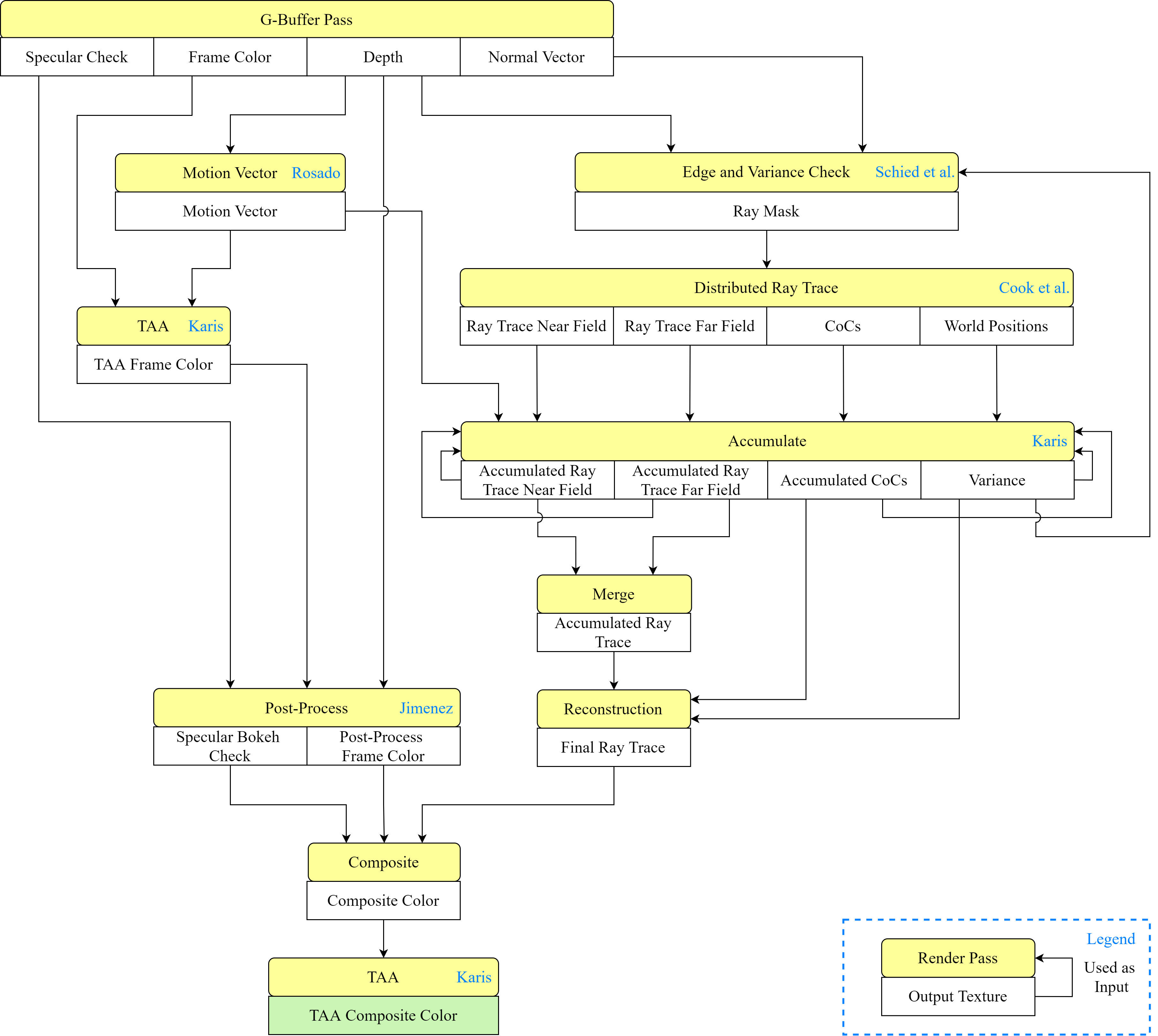}
    \caption{Hybrid rendering pipeline for DoF.}
    \label{fig:dof-pipeline}
\end{figure*}


Our approach in \autoref{fig:dof-pipeline} combines post-process based DoF with temporally-accumulated and spatially-reconstructed ray trace based DoF, to produce a hybrid DoF effect that recreates accurate semi-transparencies. Using deferred shading, a Geometry Buffer (G-Buffer) is first produced, together with textures containing other derived information needed for the post-process and ray trace stages. A sharp all-in-focus rasterized image of the scene is also generated. This image subsequently undergoes post-process filtering while parts of the scene deemed inaccurate with post-processing undergo distributed ray tracing augmented with spatio-temporal reconstruction. The images are finally composited together with a temporal anti-aliasing (TAA) pass.

\begin{figure}[!h]
    \centering
    \subcaptionbox{Near field}{
        \includegraphics[width=0.47\linewidth]{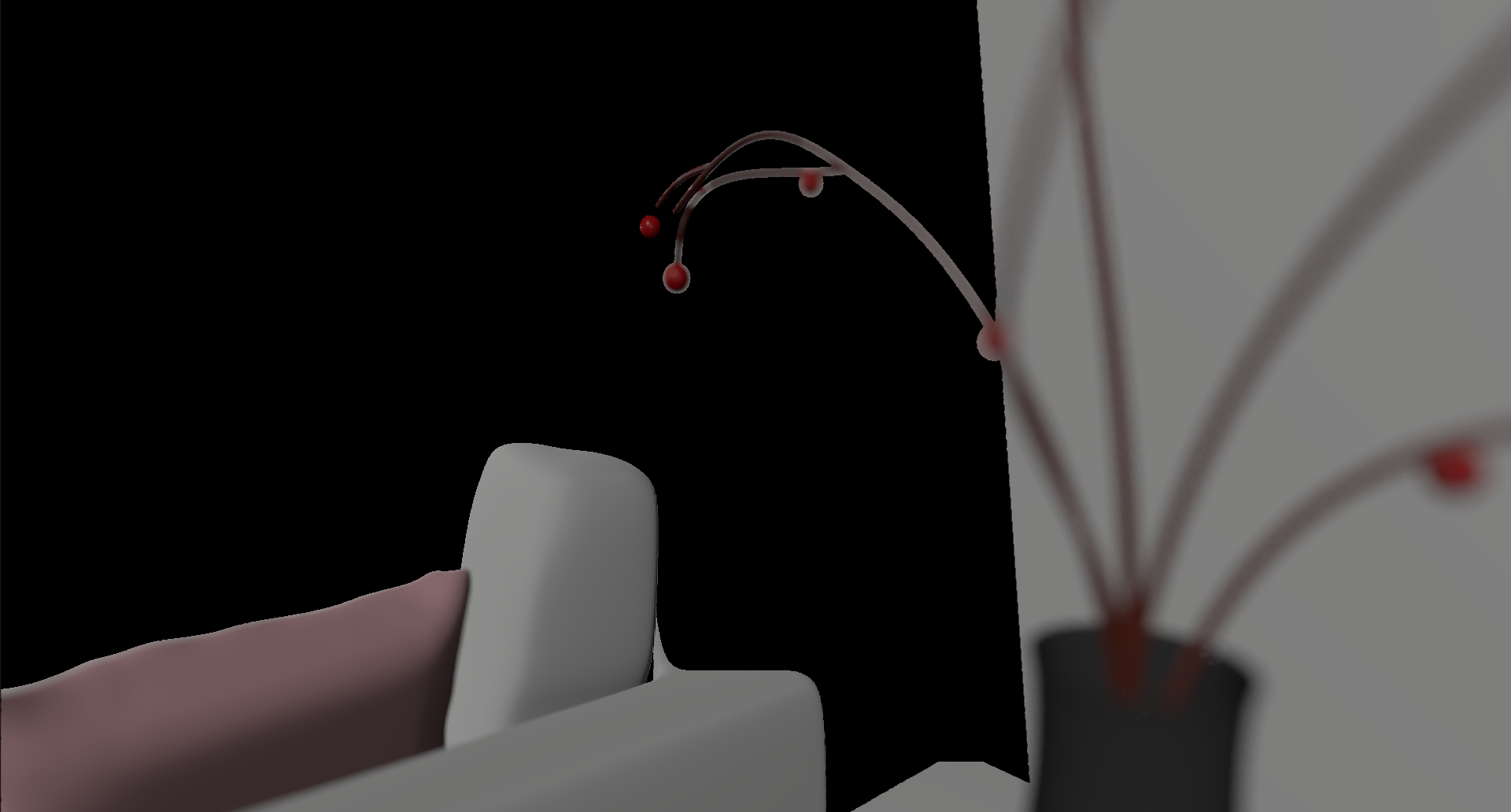}
    }
    \subcaptionbox{Far field}{
        \includegraphics[width=0.47\linewidth]{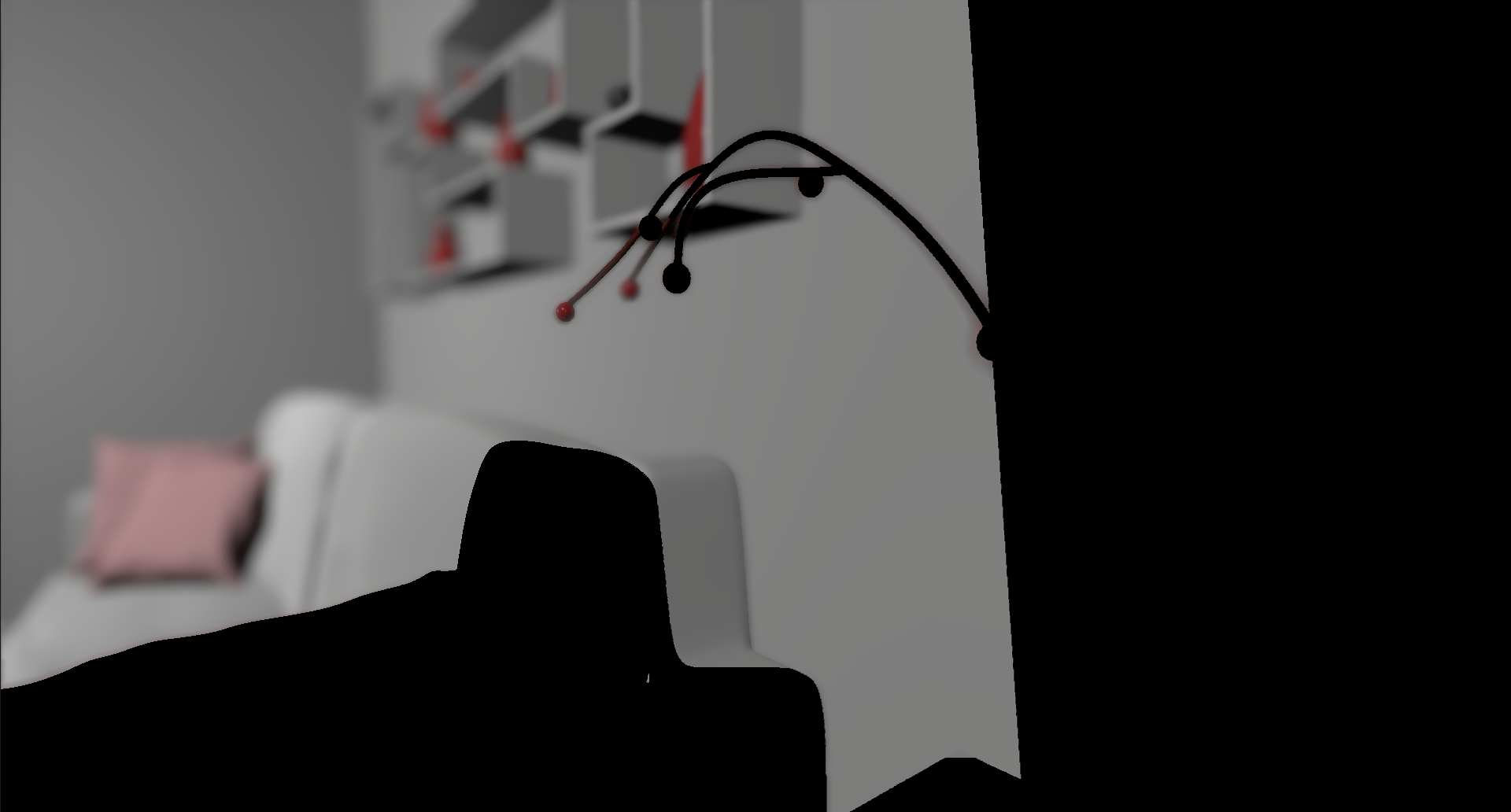}
    }
    \par\smallskip
    \caption{Demarcation of near and far fields.}
    \label{fig:dof-nearfarfield}
\end{figure}

We split our scene into the near field and the far field. Points in the scene in front of the focus plane are in the near field, and points behind the focus plane are in the far field further away, as shown in \autoref{fig:dof-nearfarfield}. We perform this split for both the post-process and ray trace images, in order to merge post-processed colour with ray trace colour on a per-field basis later on.

\subsection{Post-Process}


For our post-process technique, we adapted the DoF implementation by \citet{Jimenez:2014:ARR} which uses a gathering approach directly inspired by \citet{Sousa:2013:GGF} for filtering to produce blur. Following \citeauthor{Jimenez:2014:ARR}, the initial rasterized image is downscaled to half its resolution to speed up the filtering process.

\subsubsection{Prefilter Pass}

\begin{figure}[!h]
    \centering
    \includegraphics[width=\linewidth]{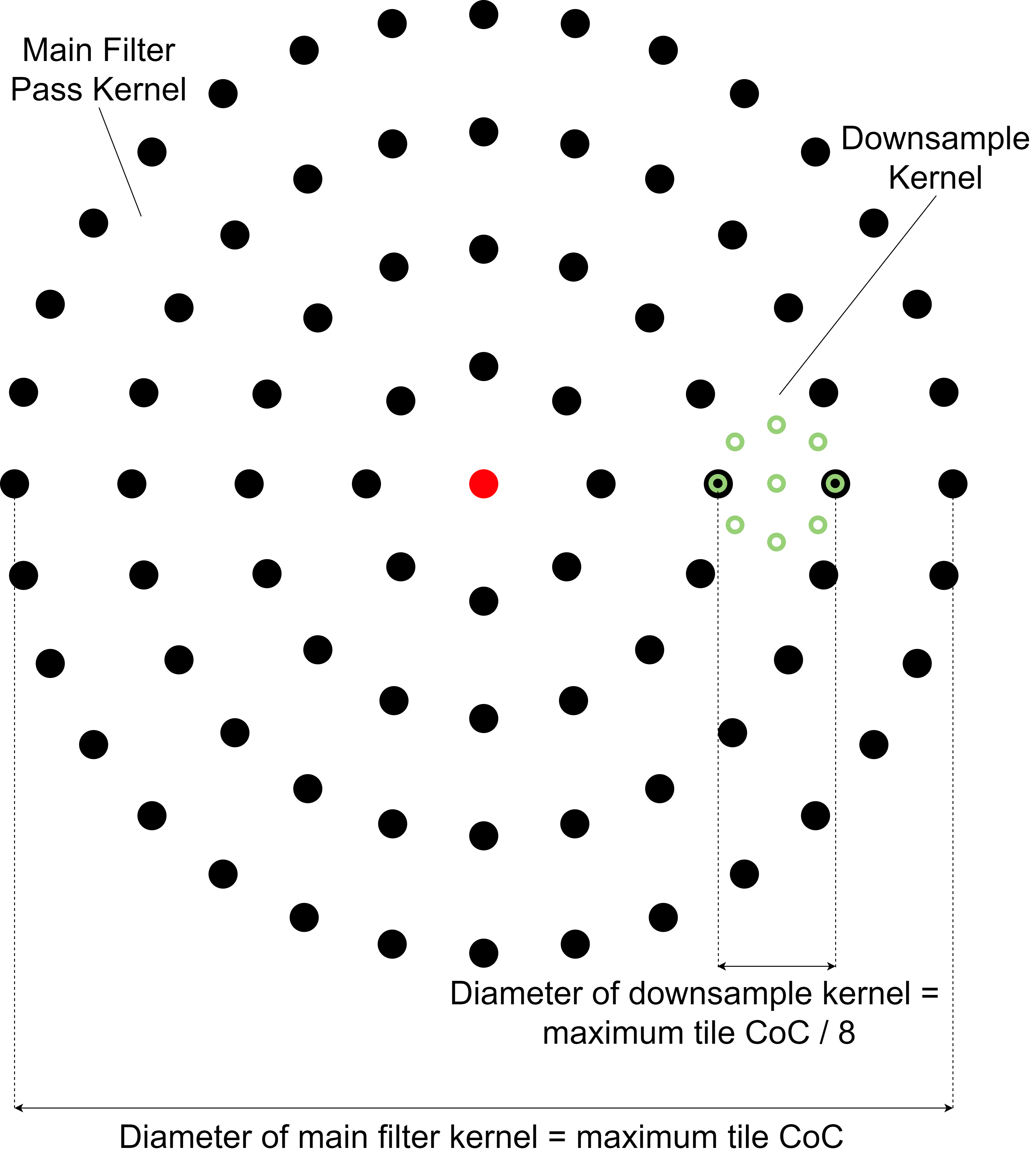}
    \caption{9-tap bilateral prefilter for 81-tap main filter.}
    \label{fig:dof-prefilter}
\end{figure}

For circular bokeh shapes, \citeauthor{Jimenez:2014:ARR} uses a 49-tap 3-ring main filter kernel scaled to the size of the maximum CoC in the tile neighbourhood of the target pixel. However, to fight undersampling, a downsampling 9-tap bilateral prefilter is first applied to fill the gaps of the main filter. We decided to use 81 taps with an additional ring of samples as shown in \autoref{fig:dof-prefilter} for better visual quality. Hence, our prefilter kernel has a diameter of $1 / 8$ instead of $1 / 6$ the maximum CoC size as in the original design. In cases where the maximum CoC is too small as most pixels in the neighbourhood are in focus, the size of the prefilter kernel is capped at $\sqrt{2}$ (diagonal length of 1 pixel) to avoid sampling the same pixel multiple times. 

\subsubsection{Main Filter Pass}

\citeauthor{Jimenez:2014:ARR} performs alpha blending on the foreground and background layers with the normalized alpha of the foreground. However, the implementation result was unsatisfactory as the normalized alpha calculated was too small, producing an overly transparent foreground. Hence, we used a normalized weighted sum of foreground and background contributions for the post-process colour $v_{p}$ instead as shown.
\begin{equation}
    v_{p} = \frac{v_{f} + v_{b}}{\sum_{i=1}^{81}\newD{$0$, $i$} \cdot \sampleAlpha{$r_{i}$}}
\end{equation}


In the above equation, $r_{i}$ refers to the CoC radius of sample $i$ while $v_{f}$ and $v_{b}$ represent the total accumulated colour for the foreground and background respectively. $\newD{$0$,$i$}$ refers to the comparison of the CoC of sample $i$ to its distance to the centre tap of the kernel. If the radius of the sample's CoC is greater than its distance to the kernel centre, the sample contributes to the target pixel's colour.

To combat aliasing, we jitter the camera's position with pseudorandom number values. We also gather the proportion of samples with high specular values for each pixel to be used to composite ray trace and post-process colour on bright bokeh shapes later on.

\subsubsection{Postfilter Pass}
\label{sec:dof-median}

Finally, as recommended by \citeauthor{Jimenez:2014:ARR}, we apply a $3 \times 3$ median postfilter at half resolution to upscale the image back to full resolution like in \citet{Sousa:2013:GGF}. The median postfilter, based on a GPU-optimized max-min network flow \citep{Smith:1996:IMF}, helps to remove noise from the main pass filtering by rejecting outlier pixels, smoothening out the result.

\subsection{Ray Trace}


\subsubsection{Ray Mask}


We shoot a variable number of rays into the scene by creating an adaptive ray mask based on the gradient of surface normals. Employing a selective rendering approach \citep{Chalmers:2006:SRC} for better performance, we aim to shoot more rays at edges to create clean semi-transparencies but less at regions with fewer details such as relatively flat surfaces. 

Our ray mask utilizes a $5 \times 5$ Sobel convolution kernel to estimate how extreme an edge is. Adopting ideas from Canny Edge Detection \citep{Canny:1986:CAE}, we apply a Gaussian filter on the G-Buffer before performing the Sobel operator so as to reduce noise and jaggies along diagonal edges. The Sobel kernel is then applied to the filtered G-Buffer at a lower resolution to get an approximate derivative of the gradient associated with each target pixel, based on the depth and surface normal of itself and surrounding pixels which are readily available from rasterization. The depth derivatives capture the separation between overlapping objects where the colour of one object might be uncovered in the other. On the other hand, normal derivatives can detect significant variations in the orientation of primitive faces within objects themselves near their silhouettes, where semi-transparencies are observed in DoF. The per-pixel output of this filter is:
\begin{equation}
    x = (\delta_{d}+ \delta_{n}) \cdot s, s \in [0, 1]
\end{equation}
\begin{equation}
    x_{n} = \saturate{$1 - \frac{1}{x + 1}$}
\end{equation}


Here, $\delta_{d}$ and $\delta_{n}$ refer to the magnitude of the derivative of depth and surface normals surrounding the pixel respectively, based on the Sobel filter. $x_{n}$ refers to the normalized $x$ and $s$ is a user-specified variable to scale down the result as it is hard to perform normalization with respect to the entire scene, resulting in the aforementioned compromise. 
    
To account for temporal variation to reduce noise in the output, we also shoot more rays at regions of high variance in luminance as inspired by \citet{Schied:2017:SVF}. Hence, the ray mask is complemented with a temporally-accumulated variance estimate ${\sigma}^2$ explained later in \autoref{sec:accumulation}. Although this variance is small, it is able to detect edges of foreground objects and specular bokeh shapes. We favour shooting more rays in these regions for a cleaner image via scaling the variance by a large weight of 100000. This amplified variance is then used with $x_{n}$ to determine the final ray count as follows.
\begin{equation}
    x_{f} = \saturate{$x_{n} + {\sigma}^2 \cdot 100000$} \cdot m
\end{equation}



\begin{figure}[!h]
    \centering
    \subcaptionbox{${\sigma}^2 \cdot 100000$}{
        \includegraphics[width=0.47\linewidth]{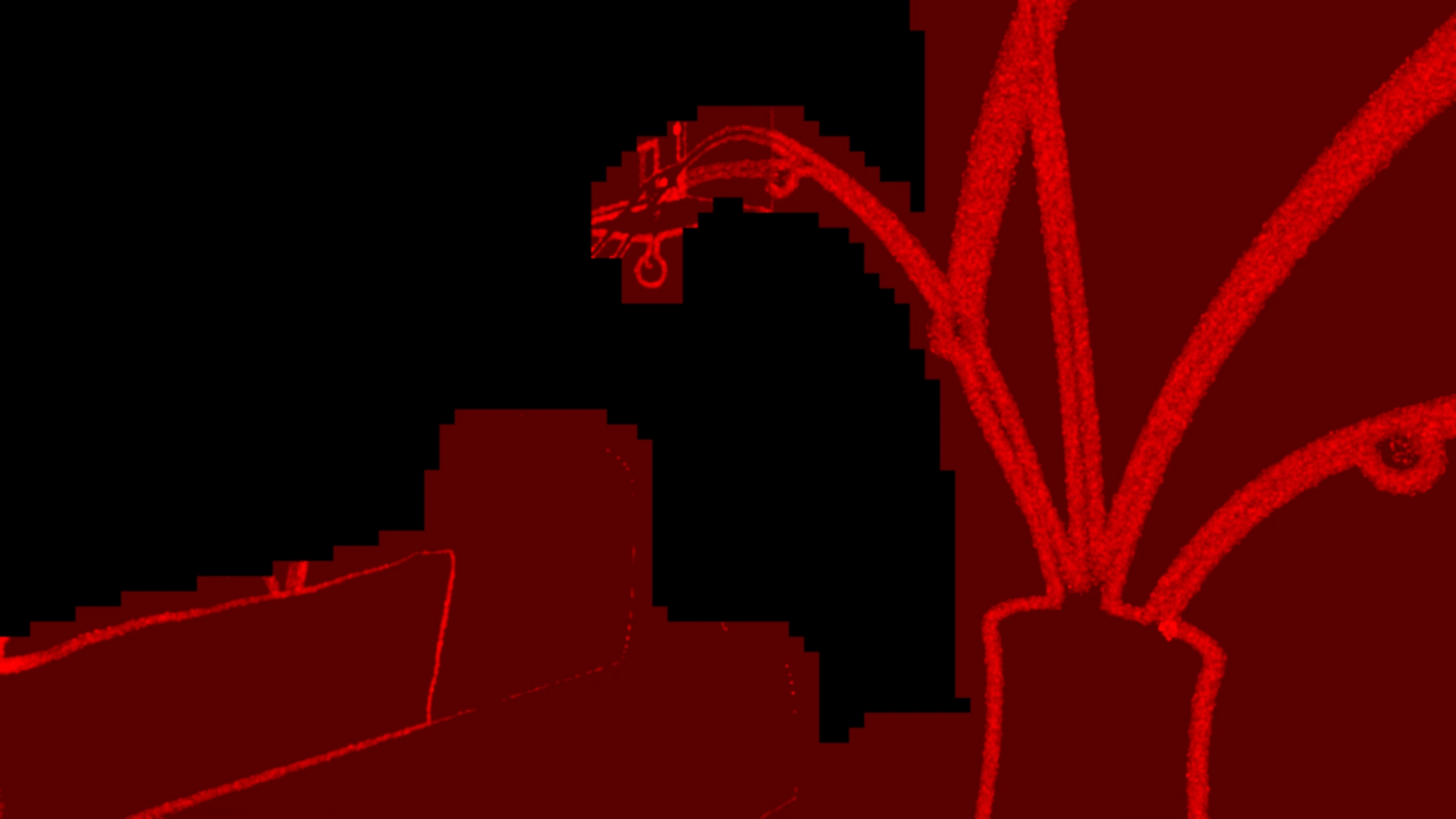}
    }
    \subcaptionbox{$x_{f}$}{
        \includegraphics[width=0.47\linewidth]{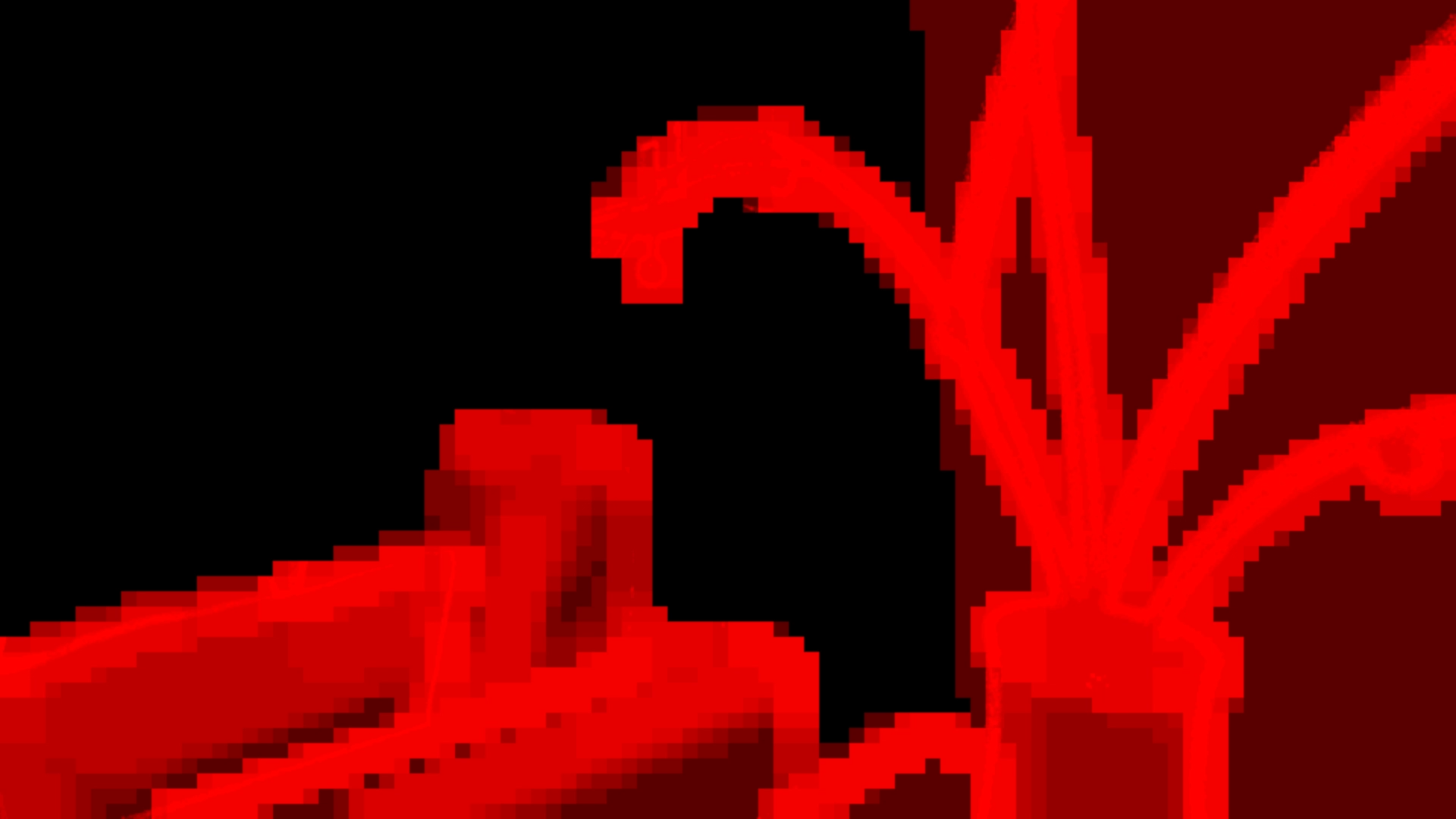}
    }
    \par\smallskip
    \caption{Comparison of variance with no. of rays shot.}
    \label{fig:dof-variance}
\end{figure}


The value of the large weight was chosen with our observation of the final ray mask generated in relation to our variance as shown in \autoref{fig:dof-variance}. Instead of selectively updating specific pixels like in \citet{Dayal:2005:AFR}, we shoot at least one ray per pixel in the near field but increase this number based on the variance gathered over time for enhanced visual quality.

\begin{figure}[!h]
    \centering
    \includegraphics[width=\linewidth]{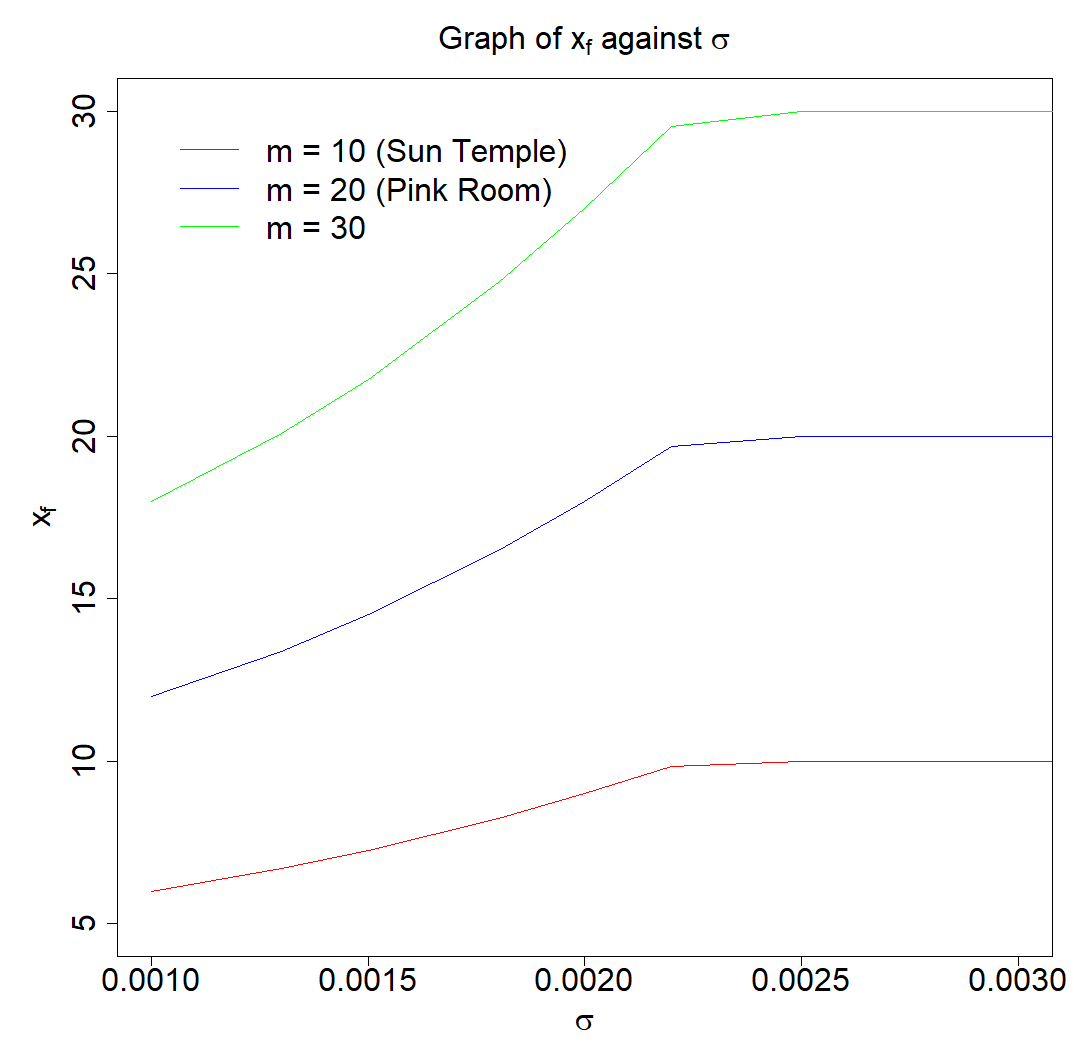}
    \caption{Graph of $x_{f}$ against $\sigma$ for different $m$ at $x_{n} = 0.5$.}
    \label{fig:dof-graphofxf}
\end{figure}

The number of rays to be shot per pixel is capped at $m$, a scene-dependent value that can be tuned for performance or accuracy. Appropriate values were chosen for scenes \textsc{Sun Temple} and \textsc{Pink Room}. As shown in \autoref{fig:dof-graphofxf}, more rays are shot as the estimated luminance variance increases. However, after a certain threshold, this number plateaus and never exceeds $m$ which is exactly the ray budget for the scene.

\subsubsection{Shooting Rays}



We follow the distributed ray tracing DoF technique by \citet{Cook:1984:DRT} in shooting a ray from a point on the lens to a calculated focus point. We consider the final image as the camera sensor and the position of the camera in the scene as the centre of the lens.

After determining the direction from a pixel on the “sensor” (the image being rendered) to the centre of the lens, we then compute the focus point of this pixel on the focus plane. Then, we take random positions on the lens and spawn rays from those positions towards the focus point. As illustrated in \citet{Kraus:2007:DOF}, we take multiple samples within a circle to produce random ray origins within the lens' area. On edges of foreground objects, some rays will hit the object while others will “go around” and sample the scene behind, yielding the effect of semi-transparent silhouettes of the blurred object in the foreground, out of the zone of focus.

We ray trace at half resolution as a trade-off for speed and upscale later using the same median filter as our post-process stage in \autoref{sec:dof-median}, helping to reduce sampling noise especially on very bright areas of foreground objects. The ray-traced colours and their respective calculated CoC sizes are then sorted by depth into the near and far field. To have a smooth transition between the near and far field, we split the contribution of each ray trace colour per pixel to each field based on its distance from the separating focus plane. A hit ratio is also stored, which is the number of rays contributing to the near field colour divided by the total rays shot. 



\subsubsection{Accumulation and Reprojection}
\label{sec:accumulation}

To increase our sample count, we use temporal accumulation adapted from TAA \citep{Karis:2014:HQT}. We accumulate the near and far field ray trace colours over time and use an exponential moving average to blend between history and current frames. By default, we use a high blend factor of 0.95 to stabilize the image. 

However, to account for movement, we leverage per-pixel depth and motion information for reprojection like in \citet{Lehtinen:2011:TLF}. As such, we use motion vectors calculated based on \citet{Rosado:2008:GGP} to reproject near field pixels. As for the far field, we require an approximation of the far field world position of our target pixel. We first attempt to compute the average world position of our target pixel from far field hits of our ray trace pass. Under low ray counts, in the event that there is no far field hit, the target pixel does not give any depth information for far field reprojection. In such cases, we obtain the average far field world position of neighbouring pixels in a $3 \times 3$ region instead. Then, we use the computed world position to calculate the previous screen space position of the target pixel. This approximates reprojection for occluded objects appearing in semi-transparent regions. 


Since reprojection is different for the near and far fields, we normalize the final colour based on the accumulated hit ratio $h$. Otherwise, we run the risk of having varying colour intensities in our merged result. Hence, we perform linear interpolation (lerp) on the near and far field colours based on $h$. We also lerp the new average near and far field CoC sizes accordingly to get an approximate CoC size for the current frame for spatial reconstruction later. However, during motion, we lerp based on the latest hit ratio rather than $h$ to prevent the ghosting (or smearing) of the far field within the silhouette of near field objects.

Like \citet{Dayal:2005:AFR}, we sample more for areas of large colour variance. Borrowing from \citet{Schied:2017:SVF}, we calculate variance estimates using luminance values of the final merged ray trace image to identify regions of high noise. This includes specular bokeh shapes that are difficult to converge as well as newly ray-traced regions. The variance texture then undergoes a Gaussian blur before it is used to determine the number of rays to be shot in the next frame.

\subsubsection{Spatial Reconstruction}

Before spatial reconstruction, the ray trace colour is median filtered (\autoref{sec:dof-median}), which helps to remove sparse unconverged bokeh shapes formed by small specular highlights, trading accuracy for image quality. For reconstruction, we use a circular kernel to gather the surrounding colour contributions of neighbouring pixels.  The kernel is scaled for sampling using the average CoC size of the current frame collected from temporal accumulation, which is also used to determine the mip level or level-of-detail (LOD) for the sampling. Samples with CoC radius smaller than their distance to the target pixel are rejected, similar to the post-process filtering approach as described earlier. Finally, we lerp the original colour and hit ratio of the target pixel with that of its neighbours by clamped variance estimates $b$ to avoid over-blurring in converged regions, as shown in the equation below.
\begin{equation}
    b = \clamp{${\sigma}^2 \cdot 2000$, $0$, $0.9$}
\end{equation}


In the above equation, we are calculating how aggressive our variable size blur is for the ray-traced output. The variance is scaled by a large weight of 2000 to blur any pixels with a small variance. Once again, this value is chosen based on observation of output as opposed to physically correct rendering. 

\subsection{Composite}

For the final image, we apply the ray trace, post-process and sharp rasterized colours onto pixels based on their $z$-distances in relation to the zone of focus. To determine the depth range of the zone of focus, we first compute the range for which the CoC size of pixels is less than $\sqrt{2}$. If the CoC of a point is smaller than a “pixel” on our camera sensor (or a sensor cell), it will appear as a single pixel in the final image and can be considered in focus. Hence, we determined that the zone of focus is the set of $z$-values where: 
\begin{equation}
    \frac{a \cdot f \cdot d}{(a \cdot f+\sqrt{2} (d-f).\frac{w_{s}}{w_{i}})} \leq z \leq \frac{a \cdot f \cdot d}{(a \cdot f-\sqrt{2} (d-f) \cdot \frac{w_{s}}{w_{i}})}
\end{equation}

Here, $a$ refers to the aperture diameter of the camera lens, $f$ its focal length, $d$ the distance between the lens and the focus plane, $w_{s}$ the sensor’s width in metric units and $w_{i}$ the image’s width in pixels.


Within the zone of focus, the full resolution unblurred rasterized colour is applied instead of the filtered ray trace or post-process colours upscaled from half resolution. Outside the zone of focus, for near field objects and their silhouettes, we then apply the ray trace colour to form accurate semi-transparent silhouettes. However, for bright bokeh shapes, we favour the post-process over the ray trace colour based on the bokeh shape intensity, i.e., the proportion of samples with high specular values gathered from the main filter pass of the post-process stage. Through this, we minimize noise and ghosting artifacts from the ray trace colour in specular bokeh shapes.


For far field geometry out of the zone of focus, we adaptively blend the ray trace colour with the post-process colour using the hit ratio. If the hit ratio is high, we favour the post-process colour as fewer rays hit the far field. On the other hand, if the hit ratio is low, this means that the number of hits in the far field is comparable to that of the near field, so more ray trace colour is used. However, there are also fewer foreground hits closer to the edges of the ray mask. Hence, when blending ray trace colour with post-process colour, using our hit ratio as-is causes blur discontinuities and tiling artifacts from our ray mask. To minimize them while trying to retain the ray-traced semi-transparencies, we only blend at hit ratios below 0.3 to produce a smooth transition between the ray trace and post-process colours, as illustrated in the formula below. If we bias towards the post-process colours ($> 0.3$), we will lose the semi-transparencies rendered by our ray trace pass.
\begin{equation}
    h = \smoothstep{$0$, $0.3$, $h$}
\end{equation}

\subsection{TAA}

Fireflies are artifacts appearing due to the sampling of very bright pixels which get spread out during spatial low-pass filtering but are not temporally stable. As the camera moves, the bright spots tend to flicker from one frame to another. This persistent flickering is likely due to the temporal instability of the initial rasterized image. Because of its low-pass filtering stages, post-processed DoF is particularly sensitive to pixel flicker and will tend to spread the small temporally unstable highlights, creating fireflies artifacts. 

Following the example from \citet{Abadie:2018:ARR} to remove flickering bright pixels, we resolve our final image with TAA \citep{Karis:2014:HQT}, reprojecting previous frames onto the newest frames. To further stabilize our result, we also apply TAA to the initial rasterized image before filtering, as well as to the image generated after filtering to stabilize camera jitter.

\section{\uppercase{Implementation}}

\subsection{Falcor}

We used the NVIDIA Falcor real-time rendering framework \citep{NVIDIA:2017:FRF} with DirectX 12 backend to utilize ray tracing acceleration and develop our hybrid DoF pipeline. We also made use of Falcor's graphics techniques library to apply Gaussian blur on our G-Buffer and luminance variance in generating our ray mask, as well as TAA to stabilize our result.

\subsection{Scene-Dependent Values}

The scenes used for testing our hybrid DoF implementation are \textsc{The Modern Living Room} \citep{Wig42:2014:MLR} and \textsc{UE4 Sun Temple} \citep{EpicGames:2017:UES}. We also tested our approach on \textsc{Amazon Lumberyard Bistro} \citep{AmazonLumberyard:2017:ALB}. 

For all 3 scenes, we used $s = 0.8$ for scaling in our ray mask as it worked well to identify edges when coupled with our variance estimation. Increasing the value would bias to shooting more rays even at relatively flat surfaces. For \textsc{The Modern Living Room} (or \textsc{Pink Room}), maximum ray values $m$ of 10 to 20 per pixel worked well. Increasing the number of rays had diminishing returns for the visual quality of bokeh shapes. For \textsc{Sun Temple} and the exterior scene of \textsc{Amazon Lumberyard Bistro} (\textsc{Bistro Exterior}), we had to keep our maximum ray count at 10 to get interactive frame rates for large ray masks. However, as \textsc{Sun Temple} and \textsc{Bistro Exterior} are more complex with textures, any noise observed was less noticeable as compared to \textsc{Pink Room}.

Regarding the depth range for our post-process main filter pass, as compared to 100 feet as suggested by \citeauthor{Jimenez:2014:ARR}, we found that a value of 10 cm worked best for our test scenes due to their relatively close geometries. As for our mip sampling in spatial reconstruction, we used a simple clamped linear mapping to determine the LOD to sample from, as illustrated in the formula below, where $c_{t}$ refers to the temporally-accumulated CoC of each pixel. This linear mapping was devised from our observation of the amount of blur in specular bokeh shapes for our test scenes.
\begin{equation}
    lod = \clamp{$c_{t} \cdot 0.05$, $0$, $3$}
\end{equation}

We hope to further simplify these scene-dependent variables for ease of use by artists. 

\section{\uppercase{Results}}

\subsection{Graphics Quality Comparison}

We evaluate our hybrid DoF method (c) against our adaptation of fully post-processed DoF \citep{Jimenez:2014:ARR} without local background reconstruction (a), Unreal Engine 4 DoF \citep{Abadie:2018:ARR} (b) and fully ray-traced DoF \citep{Cook:1984:DRT} (d) in this section as well as our demo video (\href{https://youtu.be/gJuVNFwpdwM}{link}).

\begin{figure}[!h]
    \centering
    \subcaptionbox{}{
        \includegraphics[width=0.218\linewidth]{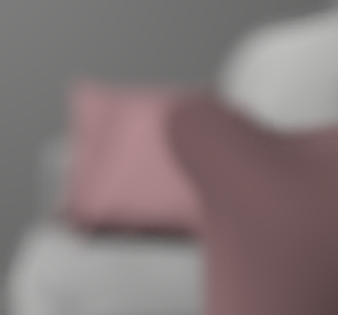}
    }
    \subcaptionbox{}{
        \includegraphics[width=0.218\linewidth]{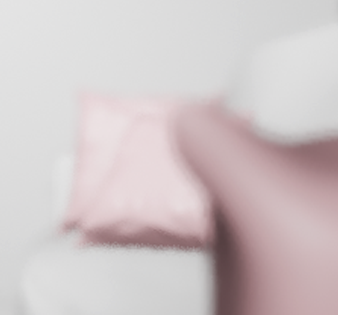}
    }
    \subcaptionbox{}{
        \includegraphics[width=0.218\linewidth]{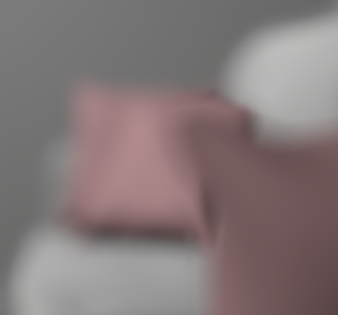}
    }
    \subcaptionbox{}{
        \includegraphics[width=0.218\linewidth]{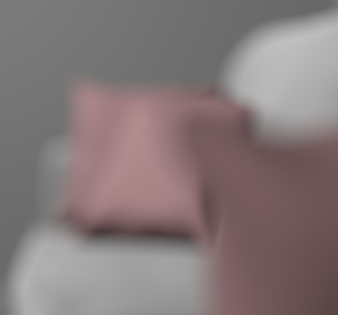}
    }
    \par\smallskip
    \caption{Semi-transparencies of foreground silhouettes.}
    \label{fig:dof-cushion}
\end{figure}

Semi-transparencies are of much better quality with hybrid DoF, as seen in \autoref{fig:dof-cushion}. For the adapted post-process DoF, the foreground blur is extended out of the object's silhouette to overwrite the background colour. However, in hybrid DoF, we can see the background object, specifically, the bottom right corner of the cushion appear along the edge of the cushion in front. This same corner is not visible for UE4.

\begin{figure}[!h]
    \centering
    \includegraphics[width=\linewidth,height=0.5\textheight,keepaspectratio]{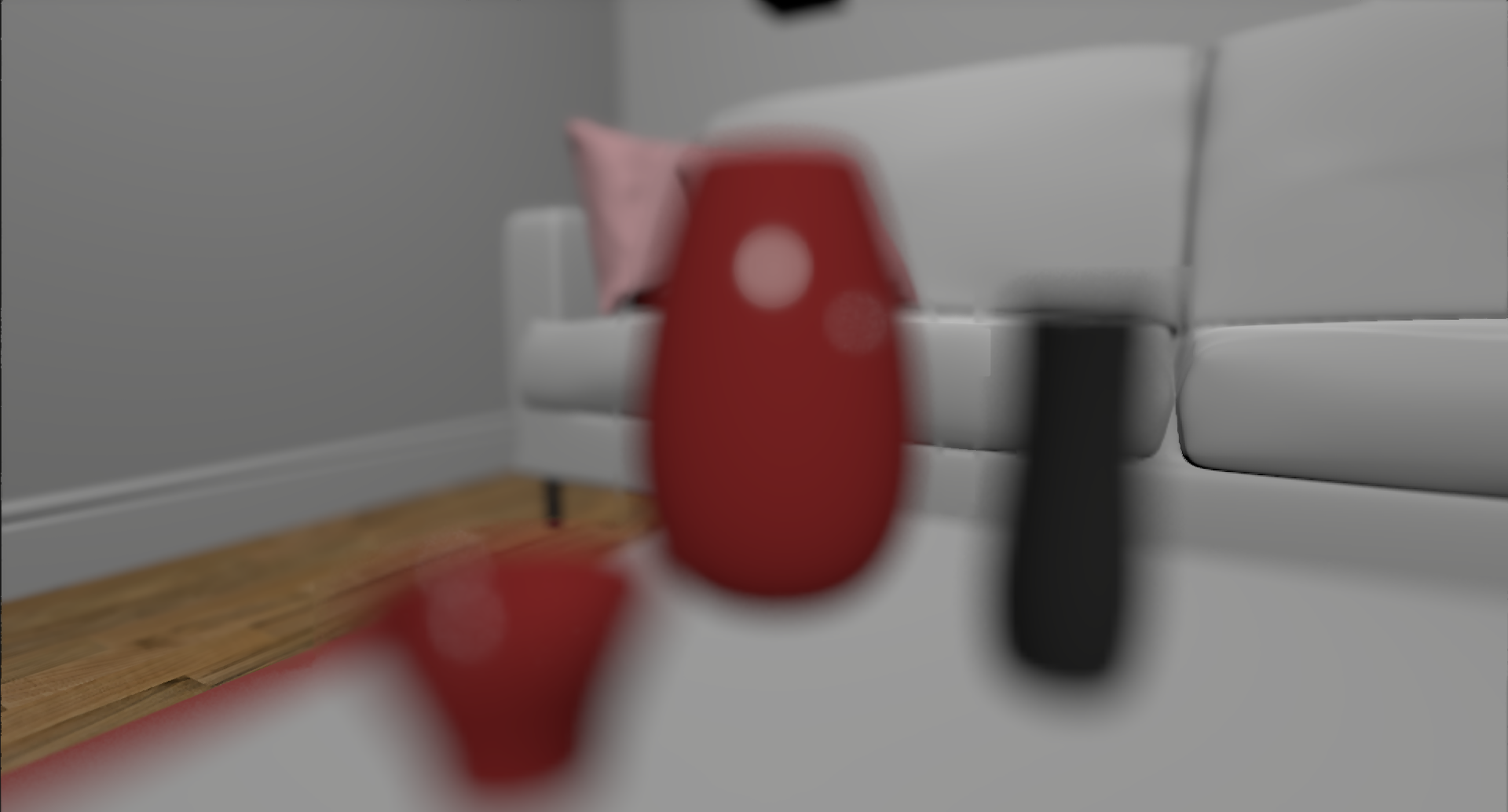}
    \includegraphics[width=\linewidth,height=0.5\textheight,keepaspectratio]{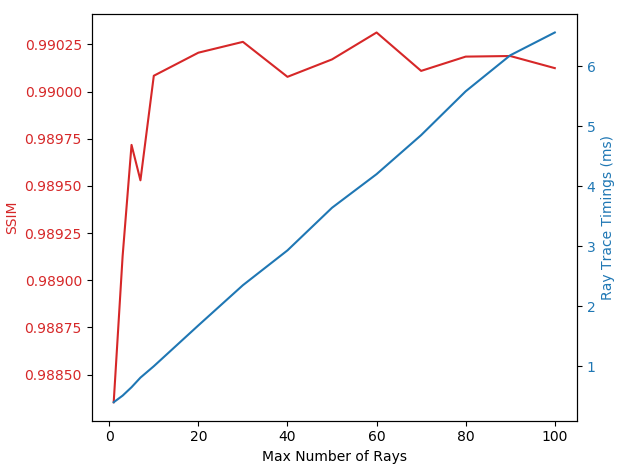}
    \caption{Top: reference image at $m$ = 10. Bottom: graph of SSIM and ray trace pass timings against ray budget $m$.}
    \label{fig:dof-ssim}
\end{figure}

To quantitatively compare the visual quality of our hybrid result with the ground truth \citep{Cook:1984:DRT}, we chose a shot of \textsc{Pink Room} with very blurred foreground geometry that exhibits highly specular bokeh shapes and semi-transparent silhouettes in \autoref{fig:dof-ssim}. When compared to ground truth DoF, increasing the maximum number of rays of hybrid DoF for the shot at relatively low ray counts increases the structural similarity index (SSIM), suggesting that our reconstruction filter might be effective in improving image quality. However, at higher ray counts, there are diminishing returns. At this stage, our technique is also unable to reach ground truth quality even at high ray counts, due to some approximated values used in our post-process stage and composition artifacts.


\subsection{Performance}

\begin{table*}[!h]
\begin{center}
    \caption{Shots used for profiling.}
    \label{tab:dof-profilingshots}
\begin{tabular}{|c|c|c|c|}
    \hline
    $m$ & 1 & 30 & 50 \\ \hline
    BG &
    \includegraphics[width=0.28\linewidth]{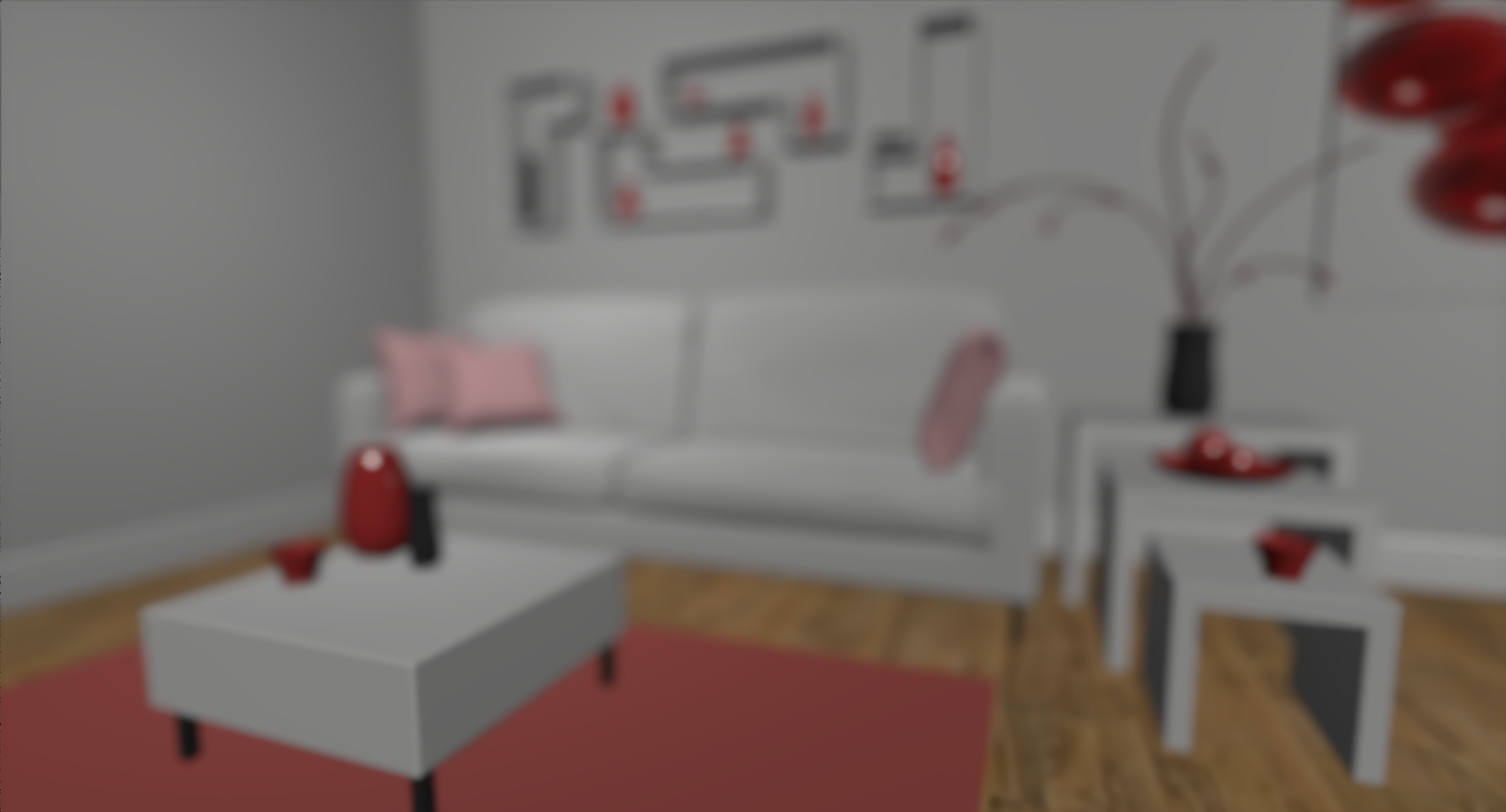} &
    \includegraphics[width=0.28\linewidth]{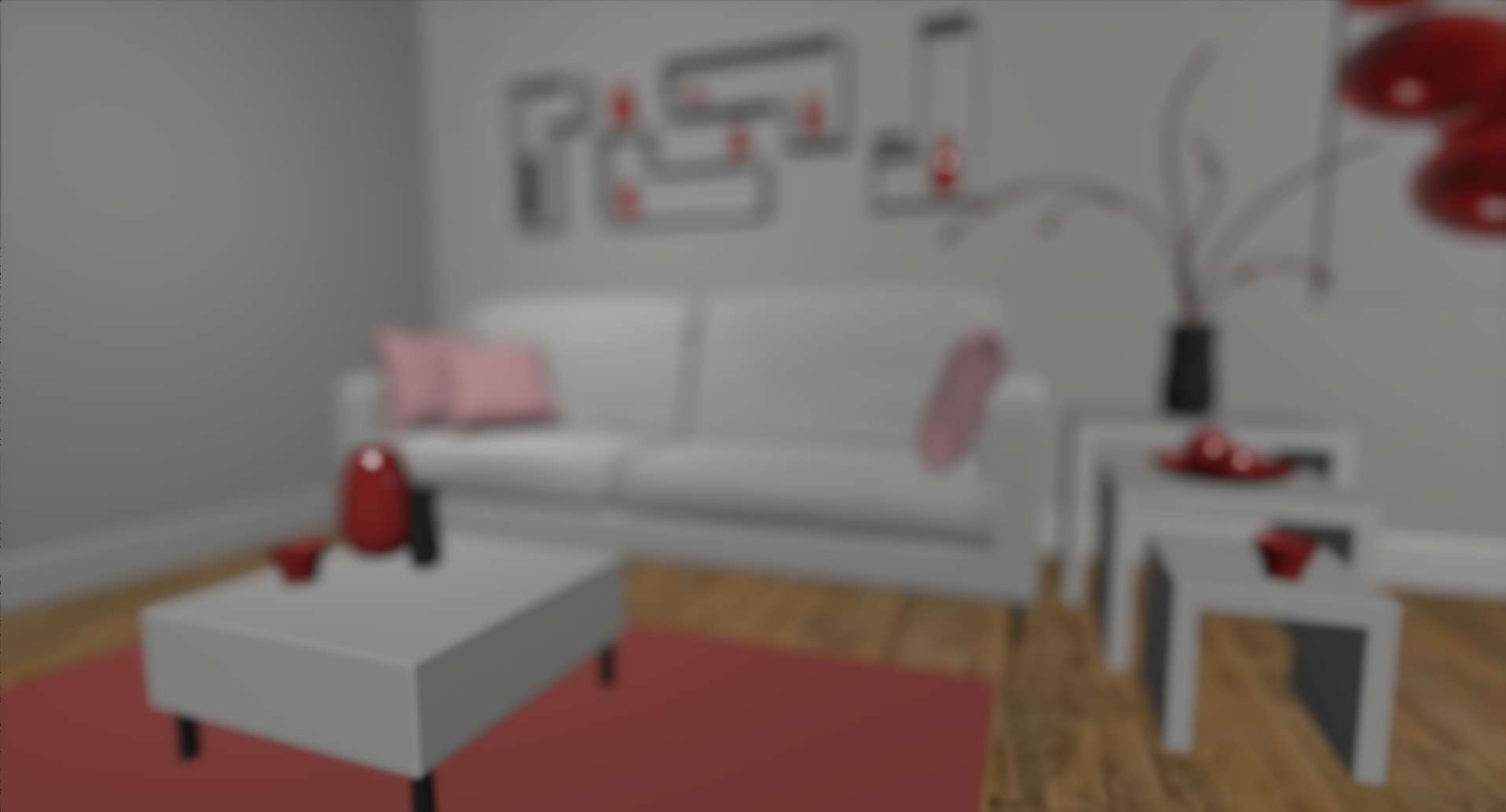} &
    \includegraphics[width=0.28\linewidth]{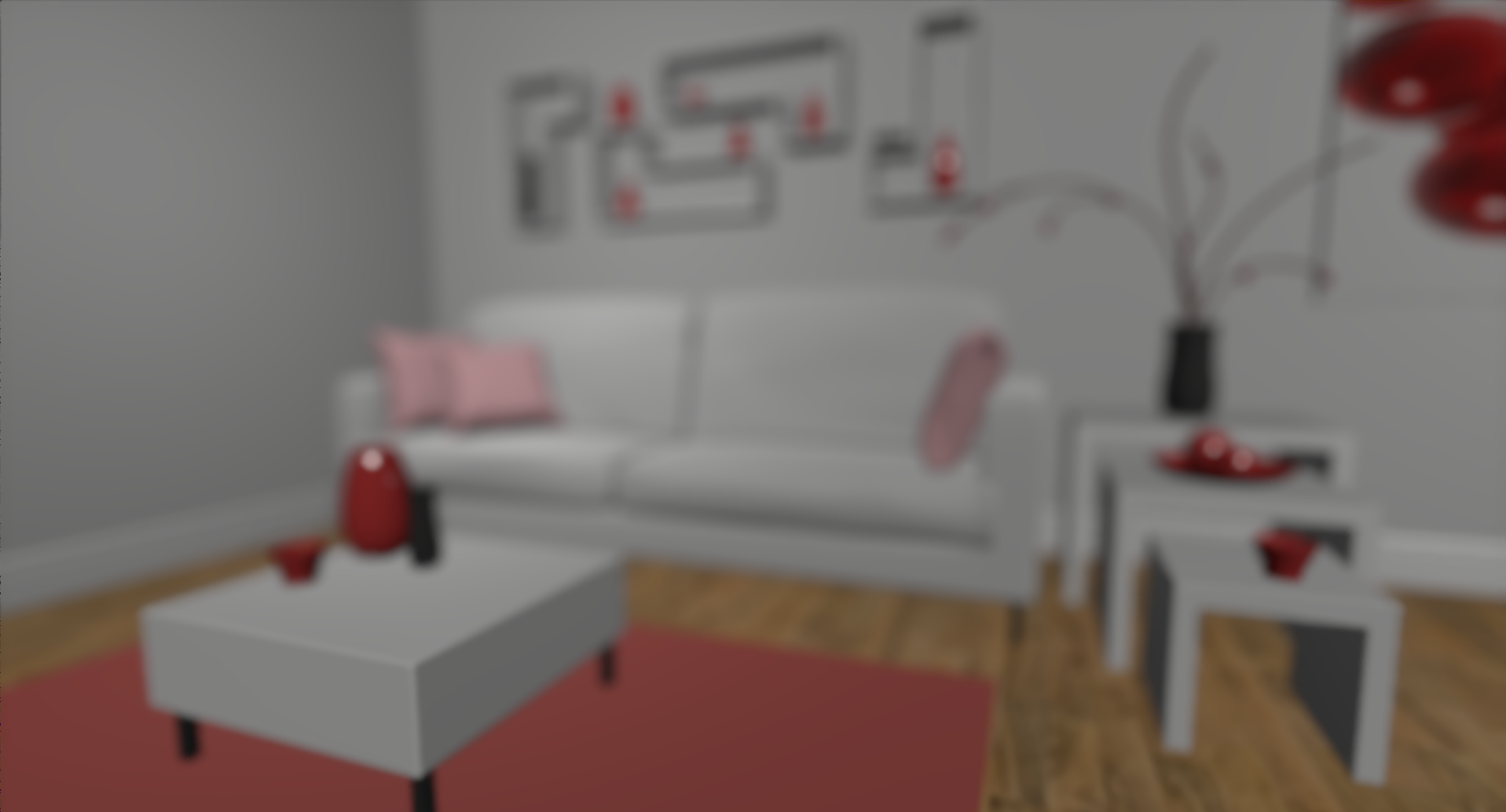} \\ \hline
    M &
    \includegraphics[width=0.28\linewidth]{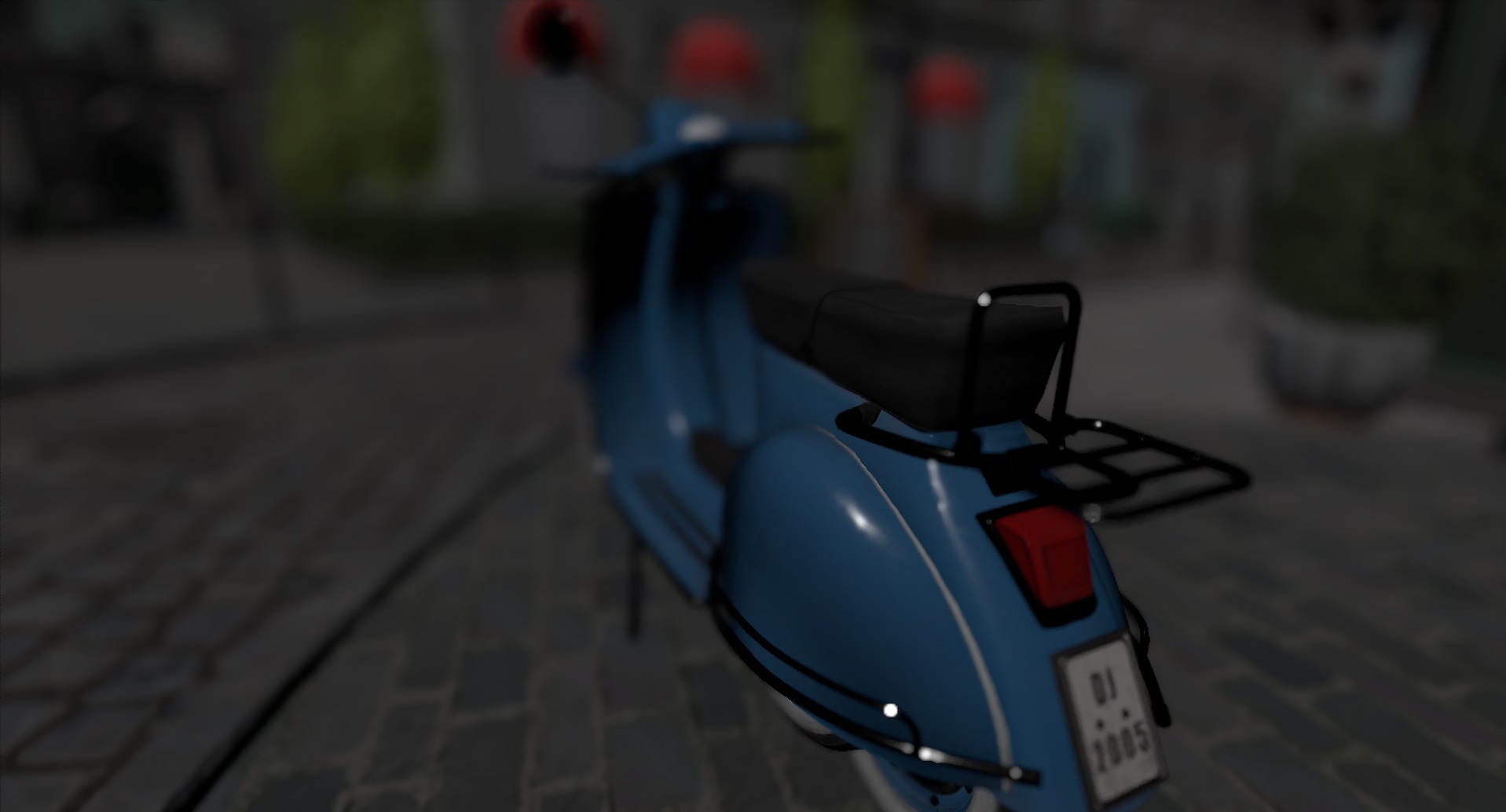} &
    \includegraphics[width=0.28\linewidth]{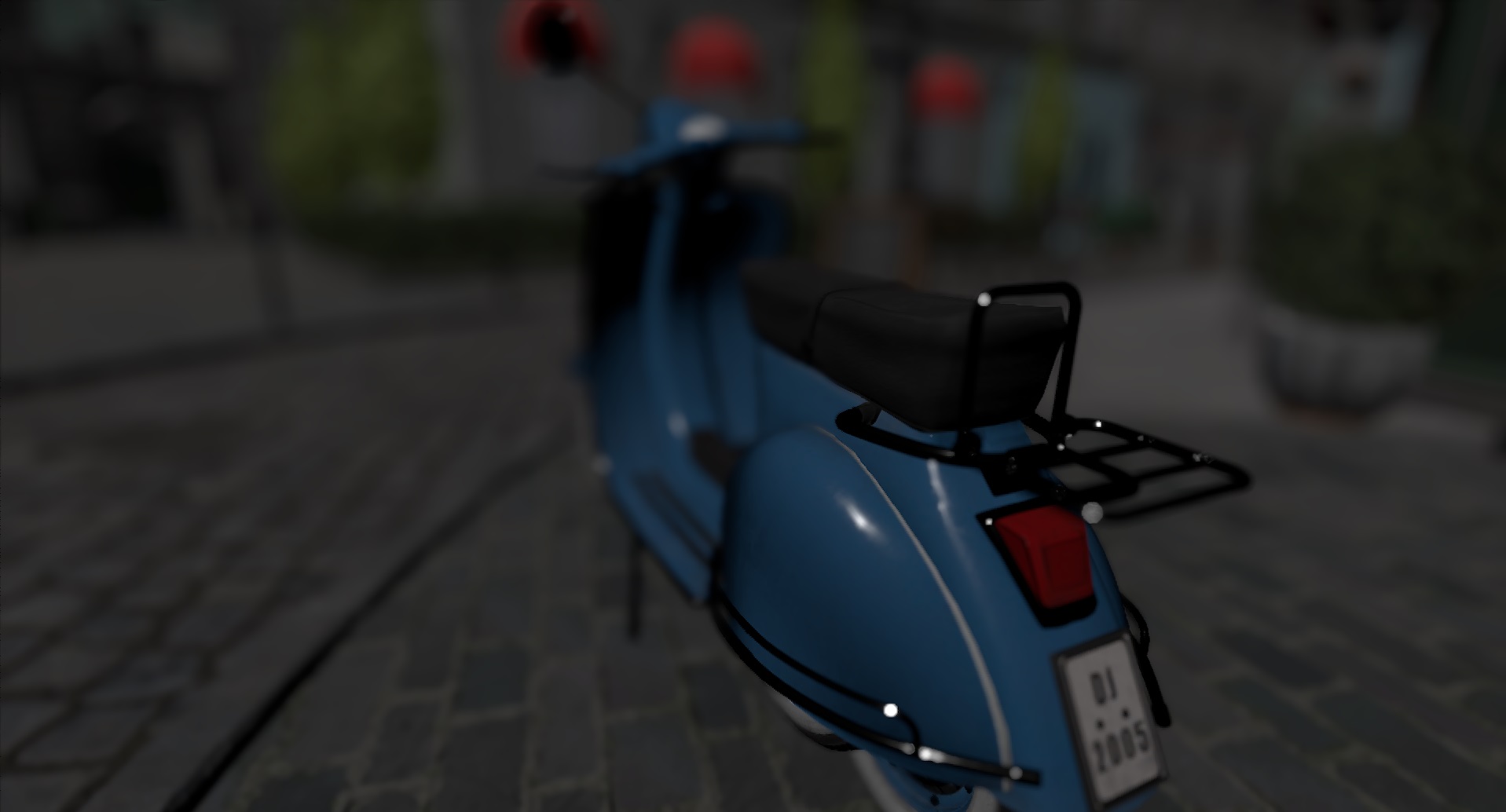} &
    \includegraphics[width=0.28\linewidth]{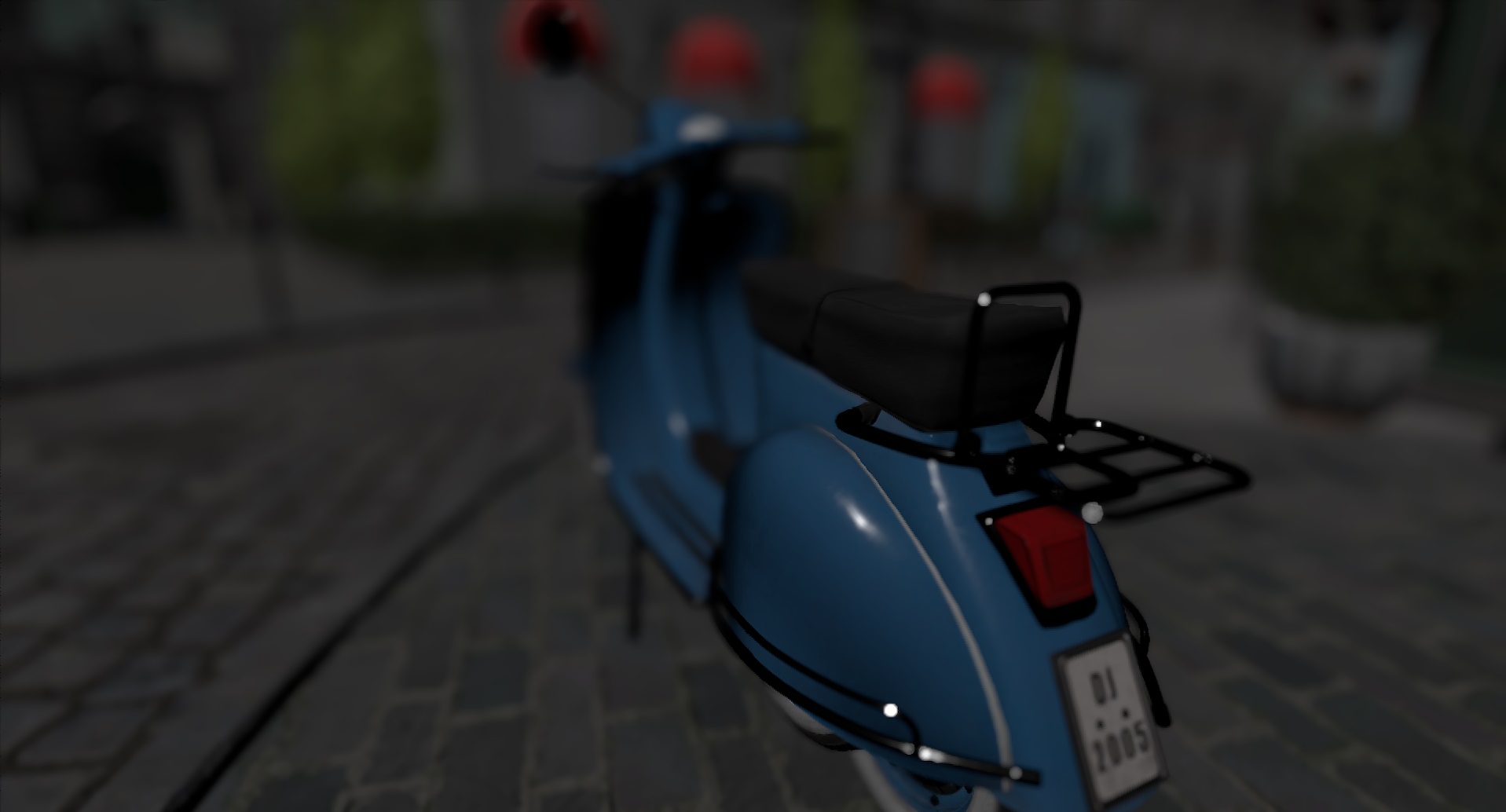} \\ \hline
    FG &
    \includegraphics[width=0.28\linewidth]{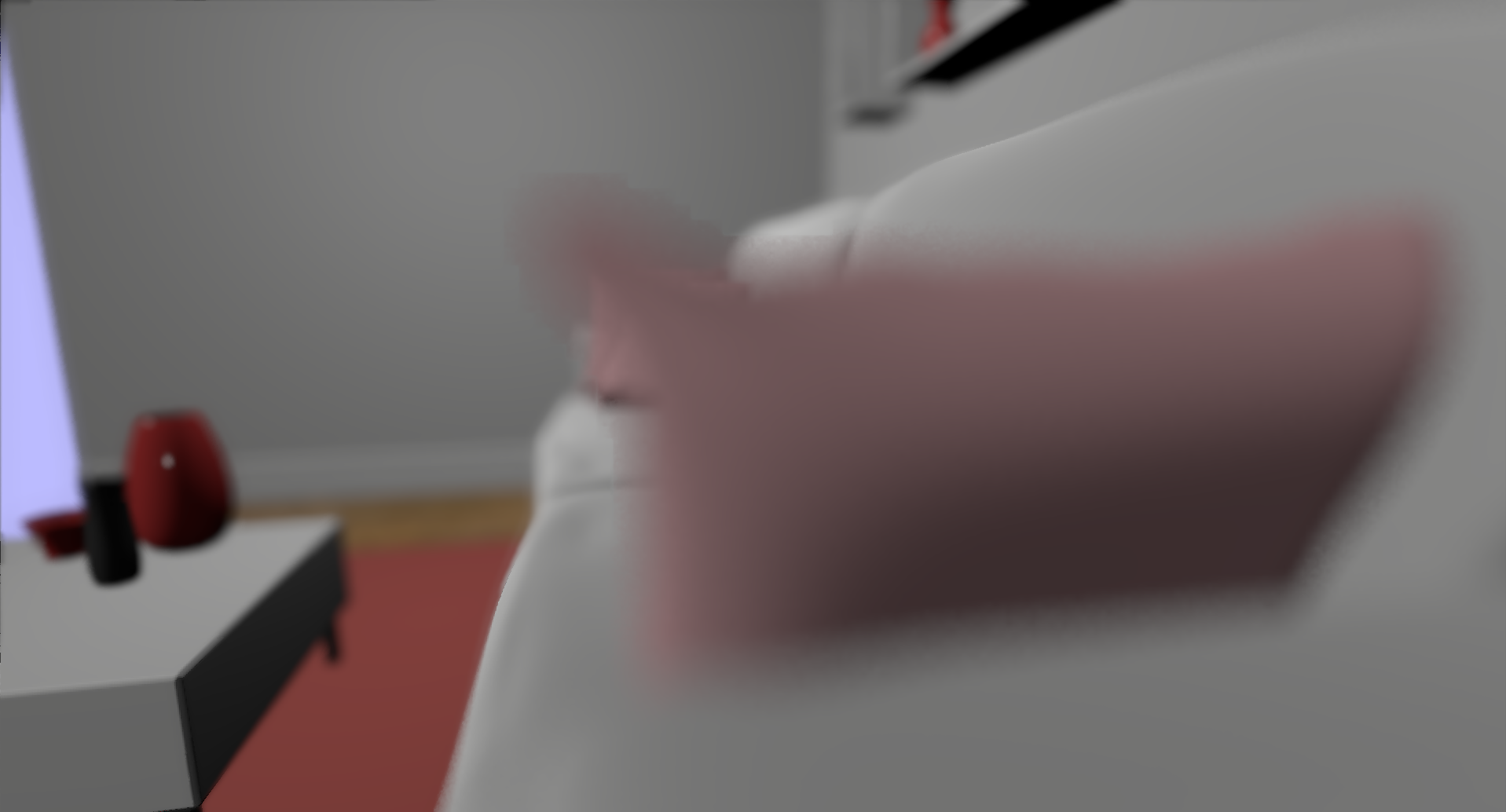} & 
    \includegraphics[width=0.28\linewidth]{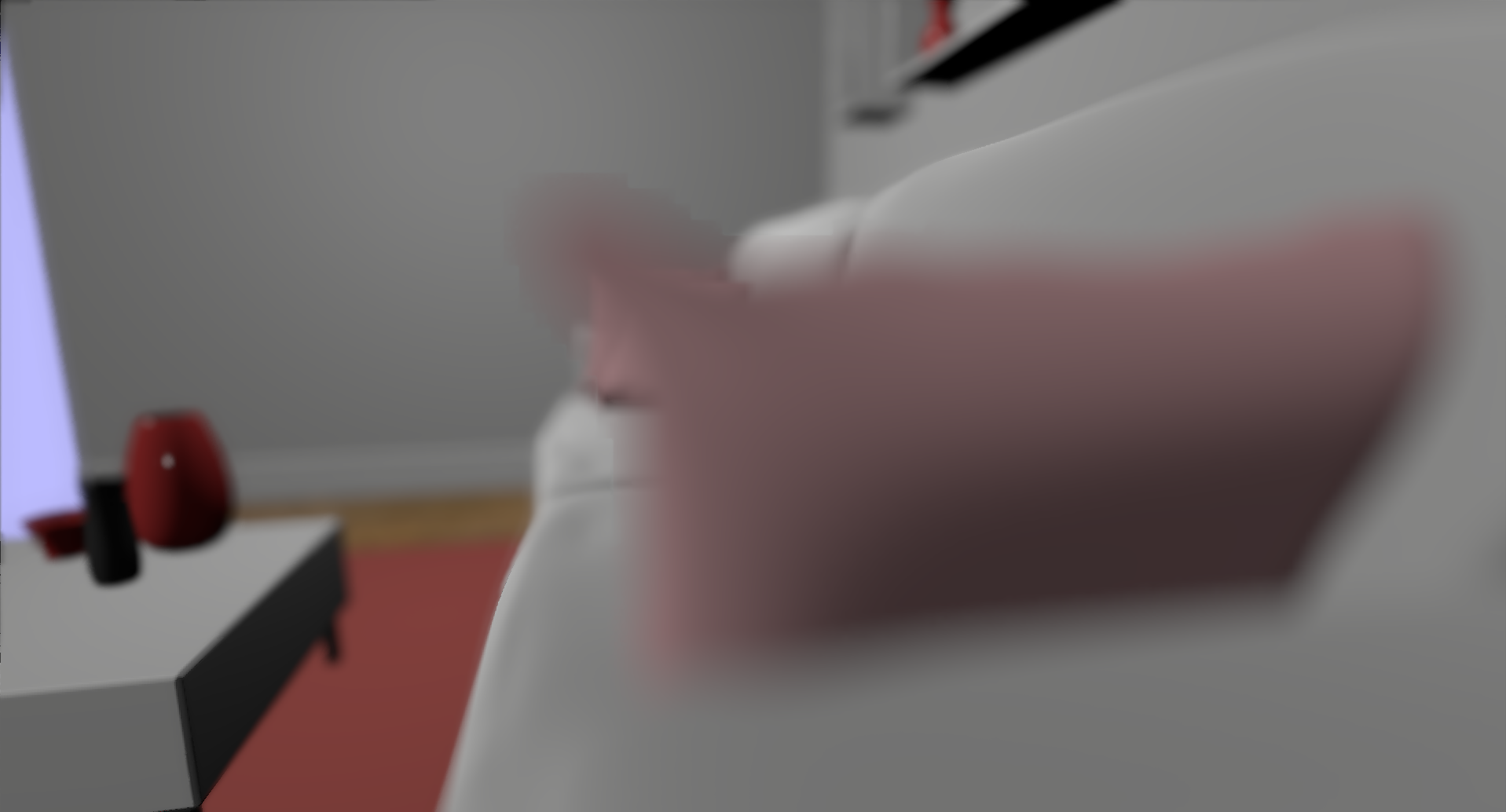} & 
    \includegraphics[width=0.28\linewidth]{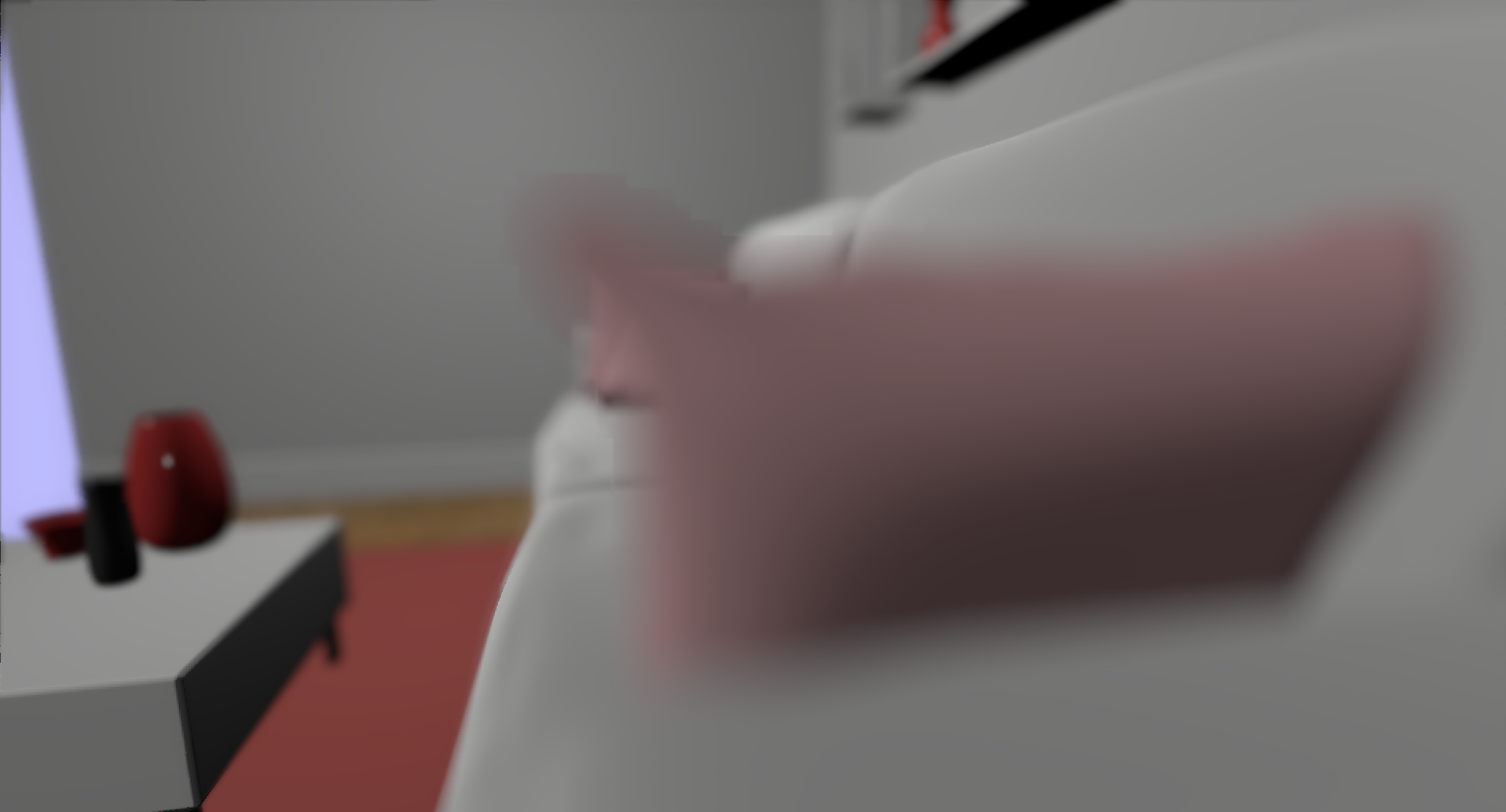} \\ \hline
\end{tabular}
\end{center}
\end{table*}

\begin{table}[!h]
    \caption{Pass durations (in ms) and frame rates.}
    \label{tab:dof-profilingmeasurements}
\begin{center}
\begin{tabular}{|c|c|c|c|c|}
    \hline
    Shot                                                 & $m$ & 1             & 30            & 50            \\ \hline
                                                         & BG  & 1.70          & 1.72          & 1.70          \\ \cline{2-5}
                                                         & M   & 2.07          & 2.05          & 2.04          \\ \cline{2-5}
    \multirow{-2}{*}{Rasterization}                      & FG  & 1.90          & 1.89          & 1.89          \\ \hline
                                                         & BG  & 1.43          & 1.44          & 1.42          \\ \cline{2-5}
                                                         & M   & 1.63          & 1.62          & 1.69          \\ \cline{2-5}
    \multirow{-2}{*}{Post-Process}                       & FG  & 1.68          & 1.67          & 1.66          \\ \hline
                                                         & BG  & 0.64          & 0.64          & 0.65          \\ \cline{2-5}
                                                         & M   & 0.66          & 0.66          & 0.66          \\ \cline{2-5}
    \multirow{-2}{*}{\textbf{G-Buffer \&} ${\sigma}^2$ \textbf{Blur}}             & FG  & 0.65          & 0.65          & 0.64          \\ \hline
                                                         & BG  & \textbf{0.18} & \textbf{0.19} & \textbf{0.18} \\ \cline{2-5}
                                                         & M   & \textbf{0.26} & \textbf{1.36} & \textbf{2.08} \\ \cline{2-5}
    \multirow{-2}{*}{\textbf{Ray Trace}}                 & FG  & \textbf{0.40} & \textbf{2.08} & \textbf{3.25} \\ \hline
                                                         & BG  & 0.60          & 0.60          & 0.60          \\ \cline{2-5}
                                                         & M   & 0.60          & 0.60          & 0.60          \\ \cline{2-5}
    \multirow{-2}{*}{\textbf{Accumulation}}              & FG  & 0.61          & 0.61          & 0.61          \\ \hline
                                                         & BG  & 0.35          & 0.34          & 0.34          \\ \cline{2-5}
                                                         & M   & 0.33          & 0.34          & 0.34          \\ \cline{2-5}
    \multirow{-2}{*}{\textbf{Median}}                    & FG  & 0.34          & 0.34          & 0.34          \\ \hline
                                                         & BG  & 0.42          & 0.42          & 0.42          \\ \cline{2-5}
                                                         & M   & 0.81          & 0.79          & 0.81          \\ \cline{2-5}
    \multirow{-2}{*}{\textbf{Recon-Composite}}           & FG  & 1.42          & 1.40          & 1.42          \\ \hline
                                                         & BG  & 0.53          & 0.53          & 0.53          \\ \cline{2-5}
                                                         & M   & 0.53          & 0.53          & 0.53          \\ \cline{2-5}
    \multirow{-2}{*}{Final TAA}                          & FG  & 0.53          & 0.53          & 0.53          \\ \hline
                                                         & BG  & 0.64          & 0.64          & 0.66          \\ \cline{2-5}
                                                         & M   & 0.64          & 0.61          & 0.65          \\ \cline{2-5}
    \multirow{-2}{*}{Others}                             & FG  & 0.64          & 0.62          & 0.65          \\ \hline
                                                         & BG  & \textbf{2.19} & \textbf{2.19} & \textbf{2.19} \\ \cline{2-5}
                                                         & M   & \textbf{2.66} & \textbf{3.75} & \textbf{4.49} \\ \cline{2-5}
    \multirow{-2}{*}{\textbf{Our Pass}}                  & FG  & \textbf{3.42} & \textbf{5.08} & \textbf{6.26} \\ \hline
                                                         & BG  & 6.49          & 6.52          & 6.50          \\ \cline{2-5}
                                                         & M   & 7.53          & 8.56          & 9.40          \\ \cline{2-5}
    \multirow{-2}{*}{Total Duration}                     & FG  & 8.17          & 9.79          & 10.99         \\ \hline
                                                         & BG  & \textbf{181}  & \textbf{179}  & \textbf{180}  \\ \cline{2-5}
                                                         & M   & \textbf{96}   & \textbf{97}   & \textbf{97}   \\ \cline{2-5}
    \multirow{-2}{*}{\textbf{Frame Rate}}                & FG  & \textbf{146}  & \textbf{122}  & \textbf{110}  \\ \hline
\end{tabular}
\end{center}
\end{table}


Using the Falcor API helped speed up implementation by handling the loading of scene assets, setting up the rendering pipeline and creating ray tracing acceleration structures. However, the API hides many low-level details from Direct3D, making it difficult to optimize rendering. Consequently, frame rate figures are not representative of how our algorithm would perform if professionally implemented and properly optimized for games. Nonetheless, we believe that our technique can be adapted and optimized for interactive rendering and motivate further research in this direction. We have achieved relatively interactive frame rates without extensive optimization, validating hybrid rendering as a proof of concept. 

Our measurements are taken with the Nsight profiling tool on an Intel Core i7-8700K CPU at 16GB RAM with an NVIDIA GeForce RTX 2080 Ti GPU. The shots used include a background-dominant (BG) wide shot and a foreground-dominant (FG) close-up of \textsc{Pink Room}, as well as a mixed (M) medium shot of \textsc{Bistro Exterior} with different $m$ as shown in \autoref{tab:dof-profilingshots}, with their performances in \autoref{tab:dof-profilingmeasurements}. In particular, the ray trace pass duration is calculated from the combined GPU timings of the Direct3D API calls for building acceleration structures and dispatching rays. 

It can be noted that post-processing performances are on par with implementations in a production engine. As a reference, Unreal Engine achieves 1.59 ms on a GTX 1080 \citep{Abadie:2018:ARR}. While the duration of other passes remains relatively constant, the cost of the ray trace pass increases with additional rays shot per pixel except for BG. This is because background geometry is predominantly rendered with post-processing, which means that few rays are traced at all. For M, part of the moped is in the foreground and is ray-traced, while the cushion and parts of the sofa are completely ray-traced for FG. Hence, an increasing $m$ gives a dip in frame rate for FG but not M. The cost of 1 ray for M and FG are 0.04 ms and 0.07 ms respectively. These observations allow us to design content-based adaptive trade-offs between quality and performance, which we defer to future work.

\subsection{Limitations and Future Work}



Although our technique improves the quality of semi-transparencies at the silhouettes of blurry foreground geometry, we acknowledge that it currently might not fare as well as other more efficient state-of-the-art post-processing approaches in the following aspects. We hope to continue working on performance optimizations alongside visual enhancements.

\begin{figure}[!h]
    \centering
    \subcaptionbox{Ghosting\label{fig:dof-bokehghosting}}{
        \includegraphics[width=0.47\linewidth]{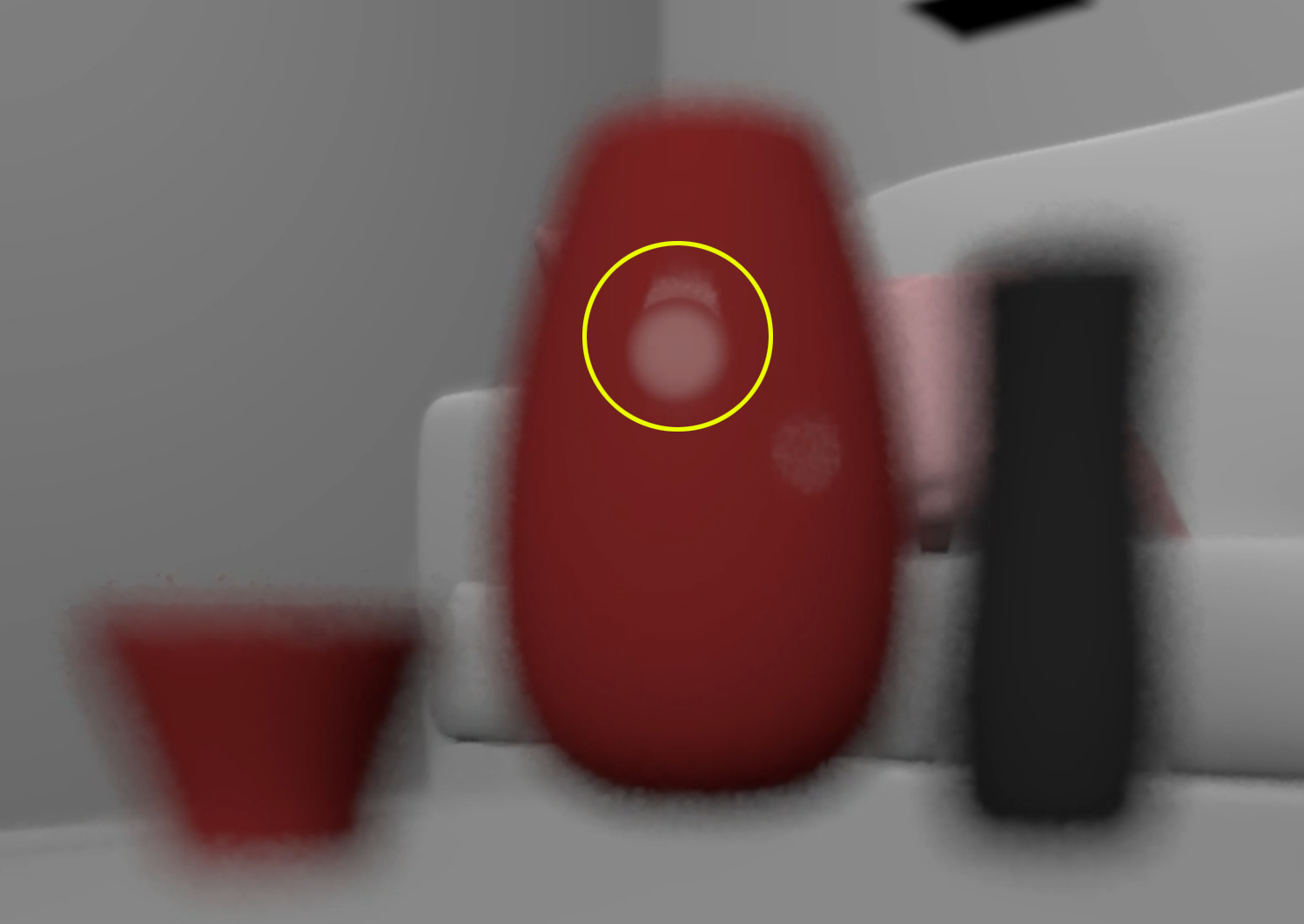}
    }
    \subcaptionbox{Tiling\label{fig:dof-tiling}}{
        \includegraphics[width=0.47\linewidth]{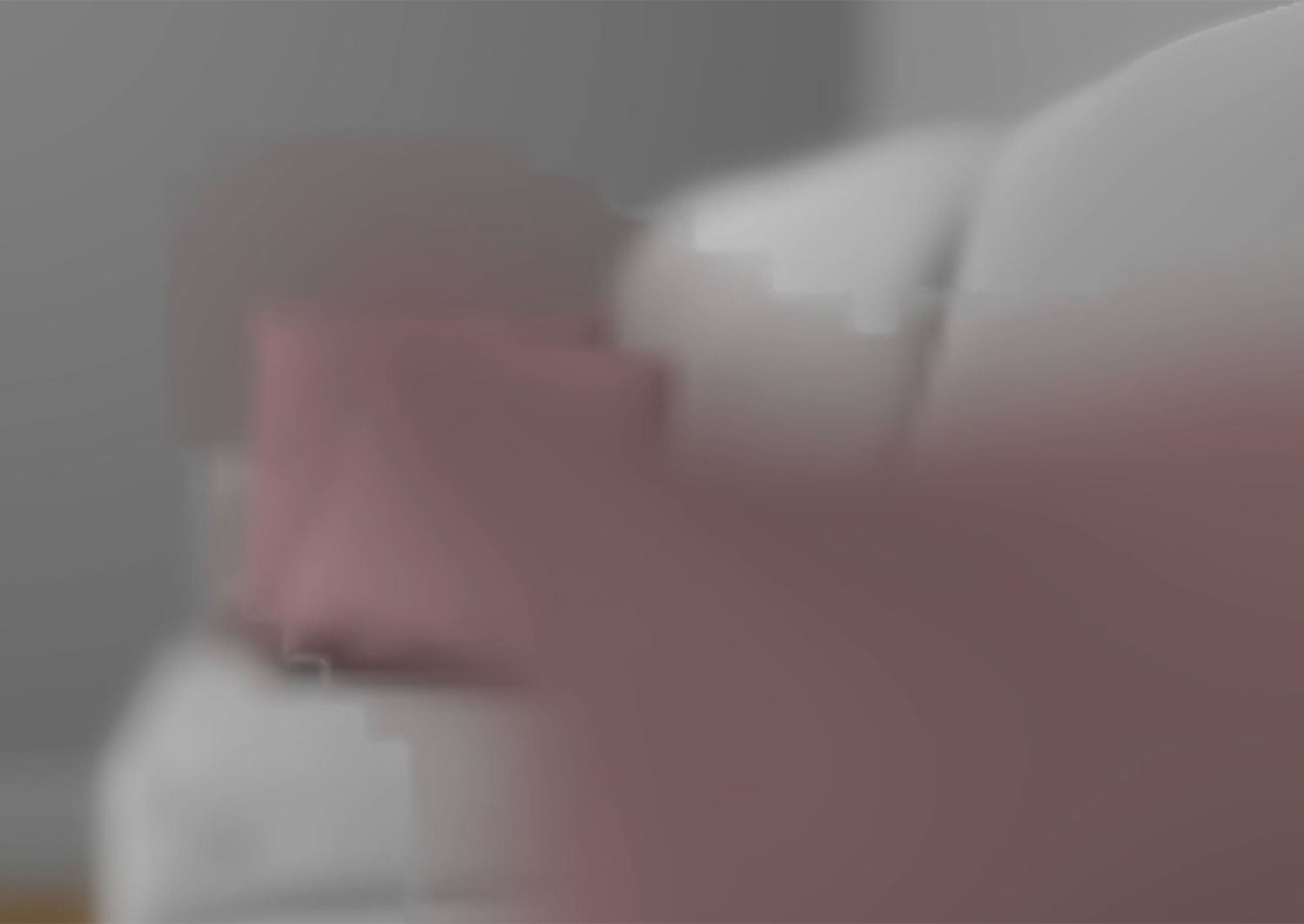}
    }
    \par\smallskip
    \caption{Artifacts.}
\end{figure}

Ghosting artifacts currently appear from temporal accumulation as shown in \autoref{fig:dof-bokehghosting}. Using colour-based neighbourhood clamping to combat ghosting as explained in \citet{Karis:2013:TMP} reintroduces noise as noted in \citet{Schied:2017:SVF}. While we have reduced the blend factor during motion, ghosting is still not eliminated especially at low frame rates as it takes longer for temporal accumulation to converge to the new colour. Potentially, we could adopt ideas from \citet{Schied:2018:GER} which manages to eliminate ghosting artifacts by estimating per-pixel blend factors.

Due to our insufficiently-sized ray mask, tiling artifacts are also observed when foreground objects are too close to the camera as seen in \autoref{fig:dof-tiling}. Ideally, our ray mask should be scaled based on CoC size to account for the sizeable blur of objects close to the camera. Hence, for our technique, we could scale our ray mask by the maximum CoC in the neighbourhood.

With 1 sample per pixel, noise generated is inherently difficult to remove. Adopting a final blur like \citet{Barre:2018:HRR} for reflections resulted in a loss of detail in semi-transparent areas. However, considering that we only ray trace within a ray mask, using a variable number of rays is a good compromise for better image quality.

As many post-process effects remain to be enhanced with ray tracing, we are also exploring hybrid rendering for motion blur. Post-processed motion blur poses similar issues of semi-transparencies. Hence, we are also investigating the use of ray tracing to uncover true background information behind motion-blurred foreground objects (\citet{Tan:2020:HMB}).




\section{\uppercase{Conclusion}}
We present a hybrid real-time rendering technique for the DoF effect in games. Our ray trace pass attains better image quality by rendering more accurate semi-transparencies with minimal artist overhead. Additionally, our ray mask and adaptive ray count, even when unoptimized, allow us to achieve relatively interactive frame rates. In future, we hope to augment and incorporate other effects like motion blur into our hybrid real-time rendering pipeline.  Our hybrid rendering engine will be open-sourced for the benefit of the research community and the industry.

\section*{\uppercase{Acknowledgements}}
We thank \citet{Wyman:2018:IDR} for the Falcor scene file of \textsc{The Modern Living Room} (\href{https://creativecommons.org/licenses/by/3.0/}{CC BY}) as well as the NVIDIA ORCA for that of \textsc{UE4 Sun Temple} (\href{https://creativecommons.org/licenses/by-nc-sa/4.0/}{CC BY-NC-SA}) and \textsc{Amazon Lumberyard Bistro} (\href{https://creativecommons.org/licenses/by/4.0/}{CC BY}). This work is supported by the Singapore Ministry of Education Academic Research grant T1 251RES1812, “Dynamic Hybrid Real-time Rendering with Hardware Accelerated Ray-tracing and Rasterization for Interactive Applications”. 

\bibliographystyle{apalike}
{\small
\bibliography{hybriddof}}

\end{document}